\documentclass[fleqn,usenatbib,useAMS,twocolumn]{mnras}
\usepackage{graphicx}    % Including figure files
\usepackage{amsmath}    % Advanced maths commands
\usepackage{amssymb}    % Extra maths symbols
\usepackage{multicol}        % Multi-column entries in tables
\usepackage{bm}        % Bold maths symbols, including upright Greek
\usepackage{pdflscape}    % Landscape pages
\usepackage[T1]{fontenc}
\usepackage{ae,aecompl}
\usepackage{newtxtext,newtxmath}
\usepackage{subfig}
\usepackage{caption}
\usepackage{multicol}
\usepackage{multirow}

\title{Optical Quasi-periodic Oscillations in the TESS light curves of three blazars}

\author[Tripathi et al.]{
Ashutosh Tripathi,$^{1,2}$\thanks{E-mail: ashutosht@tamu.edu}
Krista Lynne Smith,$^{1,2}$
Paul J. Wiita,$^{3}$
and Robert V. Wagoner$^{4}$
\\
$^{1}$Department of Physics, Southern Methodist University, 3215 Daniel Avenue,
Dallas, Texas 75205, USA\\
$^{2}$George P. and Cynthia Woods Mitchell Institute for Fundamental Physics and Astronomy, Texas A\&M University, College Station, TX 77843-4242, USA \\
$^{3}$Department of Physics, The College of New Jersey, 2000 Pennington Rd., Ewing, New Jersey 08628-0718, USA\\
$^{4}$Department of Physics and KIPAC, Stanford University, Stanford, California 94305, USA
}

\date{Last updated 2023 September 5; in original form 2023 September 5}

\pubyear{2023}
\begin{document}
\label{firstpage}
\pagerange{\pageref{firstpage}--\pageref{lastpage}}
\maketitle

\begin{abstract}
We report the time series analysis of \textsl{TESS} light curves of three blazars, BL Lacertae, 1RXS J111741.0+254858, and 1RXS J004519.6+212735, obtained using a customized approach for extracting AGN light curves. We find tentative evidence for quasi-periodic oscillations (QPOs) in these light curves that range from 2 to 6 days. Two methods of analysis are used for assessing their significance: generalized Lomb-Scargle periodograms and weighted wavelet Z-transforms. The different approaches of these methods together ensure a  robust measurement of the significance of the claimed periodicities. We can attribute the apparent QPOs to the kink instability model which postulates that the observed QPOs are related to the temporal growth of kinks  in the magnetized relativistic jet. We confirm the application of this model to BL Lacertae
%that was recently demonstrated by \cite{2022Natur.609..265J}
and extend the kink instability model to  the other two BL Lac objects.
\end{abstract}


\section{Introduction}\label{sec:intro}
%Active Galactic Nuclei (AGNs) are among the brightest non-transient objects in the Universe believed to possess supermassive black hole at its center. Objects of such high luminosity ($10^{47}$ ergs $^{-1}$) are thought to be powered by the accretion from its surroundings onto the supermassive black hole. It is widely accepted that the source of such enormous power inherited by AGN is the gravitational torques and other similar  processes. However, this hypothesis is well understood only on the larger scales ,i.e., on the order of kilo-parsecs. In reality, the accretion onto the black hole takes place on relatively smaller scales which is too small to image directly at very large distances. In order to study these processes occurring at  these smaller timescales, studying the variability of AGNs in optical waveband would be advantageous. The accretion matter around the black hole is assumed to be in local thermal equilibrium and forms disk which is optically thick and geometrically thin \citep{1973A&A....24..337S}. This accretion disk thermal emission in optical waveband. The optical variability is observed on the timescales of the order of hours to days which suggests that the emission comes from the inner most regions of the source.

Active Galactic Nuclei (AGNs) are among the brightest non-transient objects in the Universe and are understood  to possess supermassive black holes (SMBHs) at their centers. Objects of such high luminosity ($\sim 10^{47}$ erg s$^{-1}$) are thought to be powered by the accretion from their surroundings onto the SMBHs. AGNs are often classified based on their optical spectroscopic properties, which depend on their inclination to our line of sight, and on the basis of their radio emission, as radio-loud and radio-quiet objects~\citep[e.g.][]{1989AJ.....98.1195K}. Radio-quiet objects are believed to have either weak radio jets emerging close to the central object and not propagating very far or a lack of radio jet emission. By contrast, blazars are radio-loud AGNs with low inclinations, and thus with jets aligned close to the line of sight which make relativistic effects significant~\citep[e.g.][]{1995PASP..107..803U}. Blazars are usually considered to be the union of BL Lacertae (BL Lac) objects, which have very weak optical and UV emission lines, and flat spectrum radio quasars (FSRQs), for which the broad emission lines that are a defining characteristic of all quasars can be seen. Blazars have a rapidly varying continuum spanning the radio to $\gamma$-ray wavebands and exhibit strong polarization. The Doppler-boosted emission enhances the  variability throughout the electromagnetic spectrum and blazars can be distinguished by these and other extreme properties ~\citep{1979Natur.277..182S, 1979ApJ...232...34B}. Much of the radio and optical variability is thought to be due to instabilities that develop in the jet \citep[e.g.][]{1985ApJ...298..114M, 2008Natur.452..966M}.

%One of the limitation of ground based optical telescopes is the inability of taking the continuous observation of more than 12 hours in night. Moreover, these optical observations are affected by seasonal gaps, poor photometric precision and irregular sampling. Transiting Exoplanet Survey Satellite (\textsl{TESS}),similar to its predecessor \textsl{Kepler} has overcome these issues with an order of magnitude improvement in precision and sampling over the optical ground telescopes functioning currently. \textsl{TESS} will cover the whole sky with 30 minute candence with the baselines varying from 27 days to 365 days.

    Quasi-periodic Oscillations (QPOs) have been observed in X-ray light curves of stellar-mass black hole and neutron star binaries~\citep[see][for a review]{2006ARA&A..44...49R}.  The QPOs discovered in these systems are classified into low- and high-frequency QPOs. Low-frequency QPOs (LFQPOs) are very common and happen at frequencies ranging from mHz to 30 Hz. The High-frequency QPOs (HFQPOs) are much rarer and are only found in certain states of hardness and luminosity, indicating that thermal disk emission is the dominant component. The associated frequencies often occur in a 3:2 ratio \citep[e.g.][]{2006ARA&A..44...49R,2016AN....337..398M}. % Although rare, QPOs have also been apparently observed in systems possessing SMBHs in multiple bands of the} electromagnetic spectrum \citep[][and references therein]{2003ApJ...585..665H,2008Natur.455..369G,2013ApJ...776L..10L, 2017ApJ...849....9Z, 2017ApJ...847....7B,2018ApJ...860L..10S,2018A&A...616L...6G, 2019MNRAS.484.5785G, 2021MNRAS.501.5997T}.
In neutron star and stellar-mass black hole systems, these features are believed to originate from the vicinity of the black hole, probably within a few gravitational radii. Thus QPOs in these sources can be used to study  strong gravity and its effects on the surrounding plasma.
Various models have been proposed to explain the origin of LFQPOs~\citep{2016AN....337..398M}, including Lens-Thirring precession \citep{2009MNRAS.397L.101I} and unstable spiral density waves \citep{1999A&A...349.1003T}.
For HFQPOs, other theoretical models could explain the periodicities; these include warped accretion disks \citep{2005PASJ...57..699K}, disk-jet coupling \citep{2004ApJ...601..414L}, diskosiesmology \citep{2001ApJ...559L..25W}, and the relativistic precession model~\citep{1999ApJ...524L..63S}. Despite all these efforts, the physical mechanisms for the origins of QPO are still debatable. Although rare, QPOs have also been apparently observed in systems possessing SMBHs %in multiple bands of the} electromagnetic spectrum
\citep[][and references therein]{2003ApJ...585..665H,2008Natur.455..369G,2013ApJ...776L..10L, 2017ApJ...849....9Z, 2017ApJ...847....7B,2018ApJ...860L..10S,2018A&A...616L...6G, 2019MNRAS.484.5785G, 2021MNRAS.501.5997T}. The frequencies of the HFQPOs are roughly 40 times greater than those of the LFQPOs, in both stellar mass and SMBH systems~\citep{2015ApJ...798L...5Z,2021ApJ...906...92S}.

QPOs have been reported in AGNs across the whole electromagnetic spectrum, from radio to $\gamma$-rays, and claimed timescales vary from a few tens of minutes to days, to weeks, and even decades. Primarily based on the timescales of periodicity, different models have been proposed to explain these QPOs. \cite{1985Natur.314..148V} reported a periodicity of 15.7 min in the radio light curve of OJ 287 taken in 1981. \cite{1985Natur.314..146C} claimed the periodicity of 27 min in the optical observation of the same object taken 2 years later which they suggested could be explained by the presence of "hotspots" in a thick accretion disk.
Quasi-periods of the order of a few days have been reported for several AGNs \citep[e.g.][]{2003ApJ...585..665H,2018ApJ...860L..10S, 2022Natur.609..265J,2023ApJ...943...53K}. Models such as diskoseismic modes, Lens-Thirring precession, and kink instabilities were employed in those papers to explain these day-like QPOs. However, the nature of such periodicities remains uncertain and thus it is important to search for more such fast putative QPOs.

On longer timescales, \cite{1996A&A...305L..17S} reported the 11.7-year periodicity in the century-long optical light curve of OJ 287. Periods on the order of a few years have also been reported in several other \citep[e.g.][]{2001A&A...377..396R, 2003A&A...402..151R,2006A&A...455..871F,2009A&A...501..455V,2014MNRAS.443...58W,2015Natur.518...74G,2017A&A...600A.132S,2022MNRAS.513.5238R,2023ApJ...943..157L}. Those papers suggested various mechanisms that might explain such year-long periodicities;  these include the presence of a binary supermassive black hole system, a plasma blob moving helically inside the jet, and precession of the jet.

The frequencies associated with the accretion disk-based oscillations are inversely proportional to the mass of the central object. This remarkable relationship has been observed in both stellar mass and SMBHs with masses spanning many orders of magnitudes from a few solar masses to billions of solar masses~\citep[e.g.][]{2004ApJ...609L..63A, 2015ApJ...798L...5Z}. This relation suggests that the origin of such periodicity may be the same for both kinds of black holes. Moreover, this relation can be used to deduce the mass of an object if periodicity is detected in its light curve. Although SMBH masses can be plausibly estimated for some FSRQs \citep[e.g.][]{2015A&A...575A..13F, 2022MNRAS.tmp.2763X}, for BL Lacs, there are very few reasonable measurements of SMBH mass and they are model dependent~\citep[e.g.][]{ 2002ApJ...579..530W, 2021MNRAS.504.4123N}.

Optical QPOs are rarely detected in ground-based surveys due to their uneven and irregular sampling. If periodicity is searched for in such data, there is a serious risk that the stochastic red noise characteristic of AGNs can either mimic or mask a quasi-periodic signal \citep{2016MNRAS.461.3145V}. The Transiting Exoplanet Survey Satellite \citep[\textsl{TESS},][]{2015ESS.....350301R}, similar to its predecessor \textsl{Kepler} \citep{2010Sci...327..977B}, has substantially overcome these issues with improvement in precision over that yielded by the modest aperture ground-based optical telescopes normally used in these studies and  with high-cadence regular sampling they cannot provide. Thanks to \textsl{TESS}, space-based regularly sampled light curves with high precision and high cadence can be produced and used for studying periodicity at higher confidence than ground-based surveys.
%Although the rapid cadence of present-day telescopes provides astonishing details about the behavior of jets, the emission mechanism for the high-energy processes is still unknown. The interpretation of the emission mechanism may include fluid instabilities between the jet spine and sheath, or between the jet and the external medium, jet-within-a-jet models, and kink instabilities, etc.
In this work, we search for QPOs in the high-cadence and regularly sampled \textsl{TESS} observations and see that we can interpret the periodicities seen in three blazars in terms of the kink instability model.

Our team has developed a specialized approach to extracting \textsl{TESS} light curves of stochastically-varying sources like AGN. \citet{2018ApJ...860L..10S} detected a 44-day low-frequency QPO in a \emph{Kepler} light curve using a preliminary version of our current approach, which has been significantly improved to avoid over-fitting of intrinsic AGN variability while handling both background light and instrumental systematics in a careful and flexible way. \cite{2020ApJ...900..137W} analyzed the multi-wavelength observations including the observation from \textsl{TESS} of BL Lacerate (BL Lac afterward) and suggested a 13-hour timescale. A few other papers also have examined the variability of individual blazars using \textsl{TESS} observations \citep{2021MNRAS.501.1100R, 2021MNRAS.504.5629R, 2023ApJ...943...53K}.
%{\color{red}***first suggested by INSERT REF, OR SUGGESTED, NOT CONFIRMED***}.

In this paper, we report candidate QPO detections in the \textsl{TESS} light curves of BL Lacertae itself, as well as in those of two other BL Lac objects: 1RXS J004519.6+212735 and 1RXS J111741.0+254858 (which has two non-contiguous \textsl{TESS} observations). In
Section~\ref{sec:red}, we describe the sample selection and our method of reducing data taken from \textsl{TESS}. In Section~\ref{sec:analysis}, we briefly explain the various methods used for QPO detection. Results of those analysis are reported in Section~\ref{sec:res} and conclusions are given in Section~\ref{sec:con}.

\begin{figure*}
\centering
\includegraphics[width=80mm,height=35mm]{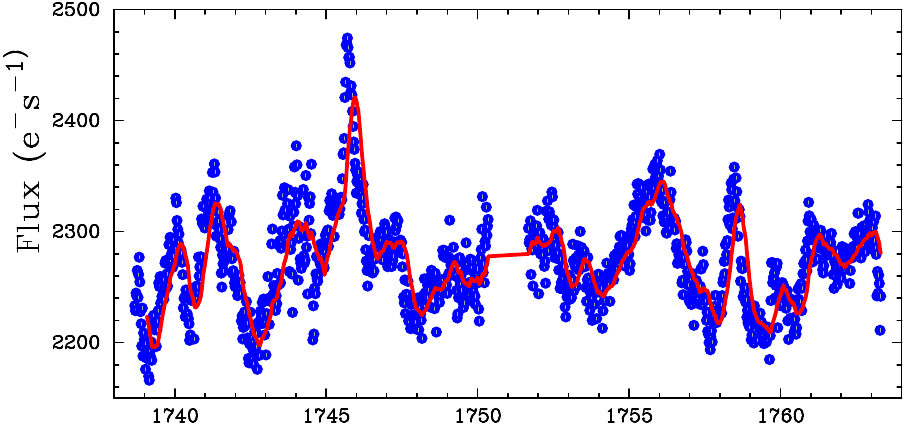}\includegraphics[width=80mm,height=35mm]{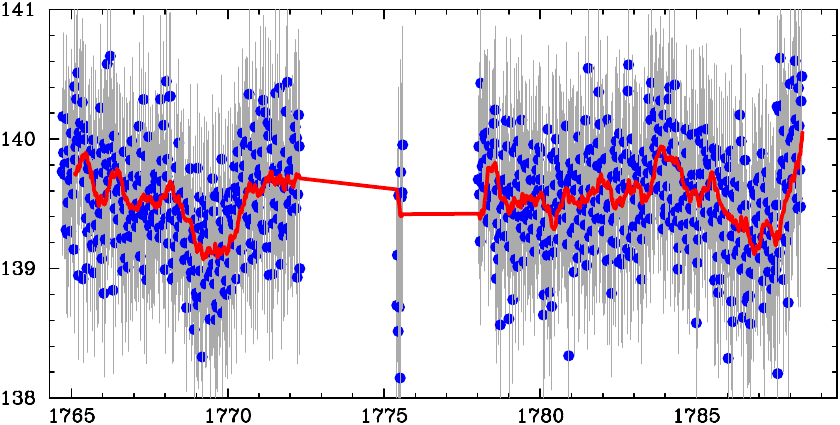}\\
%\vspace{-2mm}
\includegraphics[width=80mm,height=35mm]{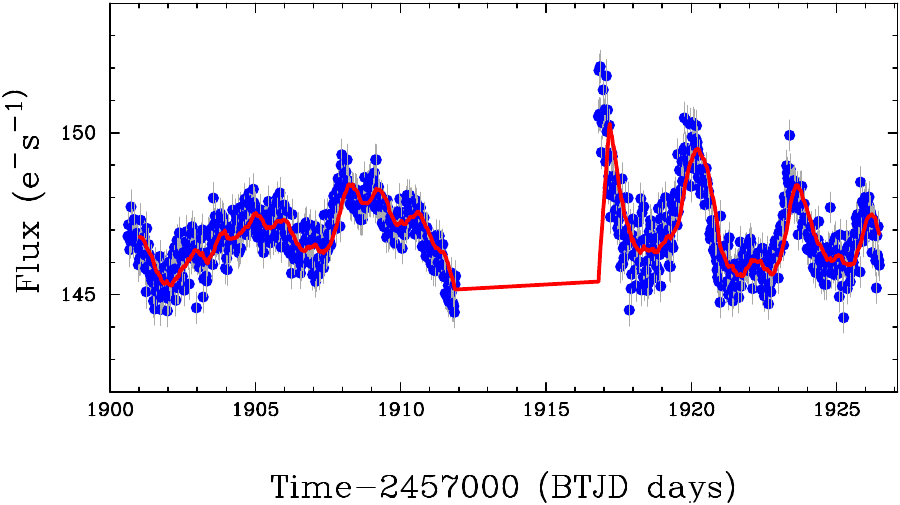}\includegraphics[width=80mm,height=35mm]{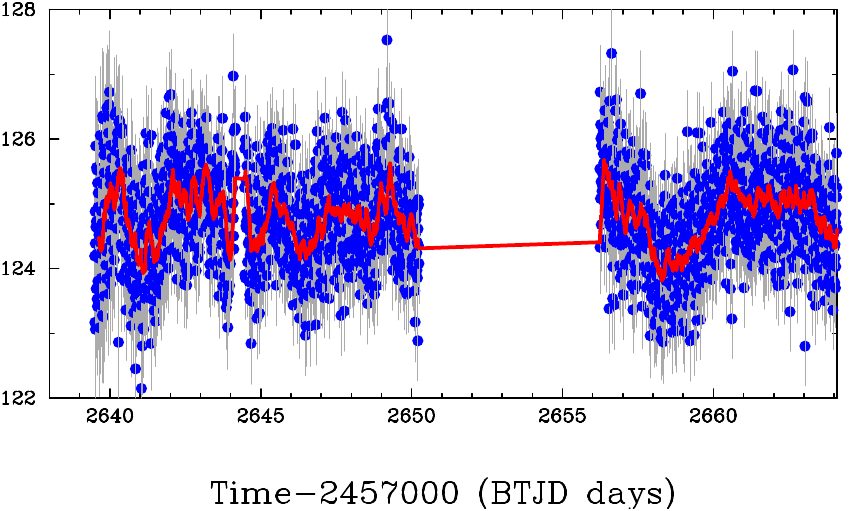}\\
\caption{\textsl{TESS} light curves analyzed in the work. The blue dots represent the fluxes and the gray curves denote the associated errors, while the red curves show the running averages. {\it Upper panel}: The left plot is of BL Lacertae and the right one is of 1RXS J004519.6+212735. {\it Lower panel}: light curves of 1RXS J111741.0+254858, where the left  and right plots respectively are for the Cycle 2 and Cycle 4 observations.}
\end{figure*}\label{fig:lc}

\section{Sample and Data Reduction}\label{sec:red}

\subsection{Sample Selection}
The candidate periodicities presented in this work were discovered  as part of an ongoing large-scale investigation of the \textsl{TESS} light curves of Seyfert galaxies. We are extracting \textsl{TESS} light curves of 5971 Type~1 Seyferts from the SDSS-IV SPIDERS catalog: a database of X-ray selected AGN from the ROSAT All-Sky Survey and the XMM-Newton Slew Survey with either eBOSS or SDSS DR12 spectral coverage \citep{2020A&A...636A..97C}. Of these objects, the preliminary periodogram analyses indicated 22 had possible periodicities. Of these, three objects survived the more rigorous test of periodic significance described below in Section~\ref{sec:analysis}. Two of them, described in this work, were BL Lac objects, and one is a Seyfert, which will be discussed in terms of the broader Seyfert sample in an upcoming work.

Because the BL Lac objects were contaminants in the SPIDERS sample of X-ray selected Seyferts, it is not possible to use these numbers to draw statistical conclusions about the occurrence rates of periodicities in blazars. In this large sample, there were only 28 blazars that exhibited detectable variations. The two presented here, alongside the \textsl{TESS} light curve of BL Lacertae itself, are the only ones showing possible indications of QPOs and are serendipitous discoveries that can provide support for the physical interpretation of the QPO in BL Lacertae reported recently by \citet{2022Natur.609..265J}.

\subsection{The Transiting Exoplanet Survey Satellite (TESS)}
\label{sec:tess}

The \textsl{TESS} instrument, designed primarily to search for transiting exoplanets using long-term monitoring at high photometric precision of stars across nearly the entire sky \citep{2015ESS.....350301R}, provides very high cadence optical monitoring, with a cadence of 30 minutes in the early cycles (2018 -- 2019) and 10 minutes or 2 minutes in later and current cycles (2020 -- present). Coverage is nearly continuous, as there are no gaps due to daylight or seasonal sunlight. \textsl{TESS} light curves are therefore significantly better-sampled than ground-based optical light curves, with longer continuous baselines. The length of nearly continuous monitoring for a \textsl{TESS} source depends upon its ecliptic latitude, ranging from 27 days nearest the ecliptic plane to approximately 1 year at the ecliptic poles. After each year-long Cycle, the spacecraft swaps hemispheres and performs the same survey in the other half of the sky. Thus, sources will occasionally have two ``Sectors'' of data, separated by one year. The mission began in Cycle~1, in the southern hemisphere, then went to the northern sky in Cycle~2, and returned to the south for Cycle~3, then the north for Cycle~4. The photometric precision of \textsl{TESS} is dependent upon the source magnitude; at the magnitudes of our sources, variability at the $\sim1-10$\% level can be confidently detected.

\subsection{TESS Data Reduction}
\label{sec:tessreduction}

\begin{figure*}
\centering
\includegraphics[width=\textwidth]{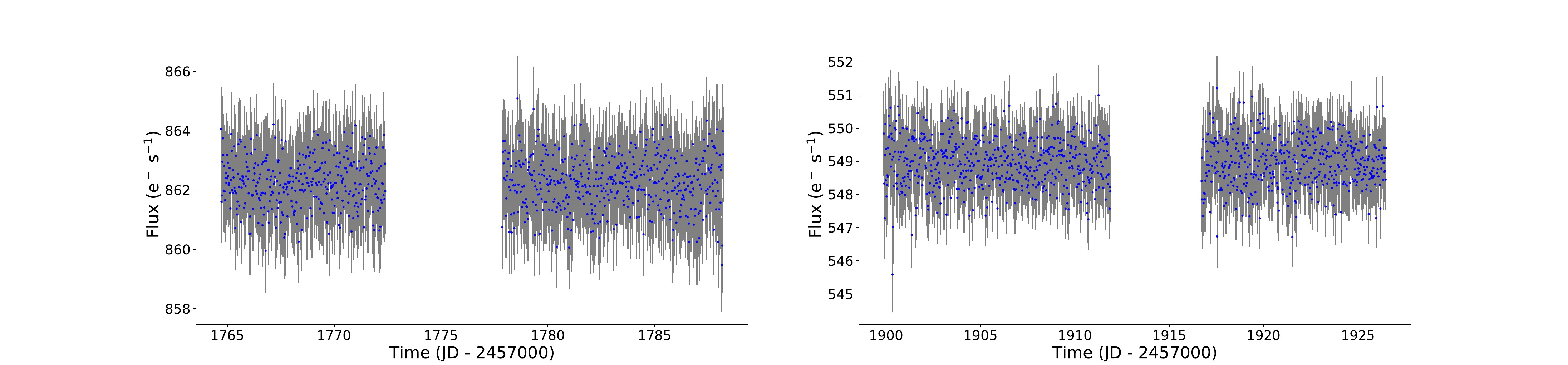}
\caption{Light curves of inactive stars obtained from same postal stamp as 1RXS J004519.6+212735 (left) and 1RXS J111741.0+254858 (right) using our customized approach.}
\end{figure*}\label{fig:sample_stars}

Because the \textsl{TESS} spacecraft was designed to detect a periodic signal in a point source, special treatment is required to derive light curves of stochastic signals that may be in resolved sources like bright host galaxies. Many of these techniques to study AGN were first developed using very similar data from the \emph{Kepler} spacecraft  \citep{2018ApJ...860L..10S}. We have developed these techniques significantly, to avoid over-fitting of intrinsic AGN variability by carefully modeling the background additive light curve effects and the subtler multiplicative systematics in separate phases using principal component analysis (PCA), then removing these effects using linear regression\footnote{\url{https://github.com/kristalynnesmith/quaver}} \citep{Smith:2023vdc}.

We first download a ``postage stamp'' of the \textsl{TESS} full-frame images around our sources, 25$\times$25 pixels on a side. From this image, we select a custom extraction aperture for each target, allowing us to avoid the inclusion of nearby sources and to build an aperture appropriate for any spatially resolved host galaxies.

Next, we define a flux threshold of 1.5~$\sigma$ above the background to determine what pixels
in the downloaded postage stamp are likely to contain only background sky or very, very faint
sources at the background level. These faint pixels are fit by three principal components. The
results of the fit are stored in a design matrix, representing the additive systematics
dominant in the light curve. This method makes use of the matrix regression methods in the
publicly-available TESSCut \citep{2019ASPC..523..397B} and  {\tt lightkurve} \citep{lightkurve}
software packages.

Next, all of the pixels in the field excluded from the previous mask, i.e., the ``bright'' pixels, but excluding the aperture of the source itself, are corrected by these additive background systematics. Once corrected, the variability of these bright pixels is again fit by three principal components, which are stored in a design matrix; this matrix describes the higher-order and more subtle multiplicative systematics. Armed with these two matrices, we combine them into a hybrid design matrix that contains principal component vectors for additive and multiplicative effects.

Finally, the raw light curve of the source is extracted from its aperture and linearly regressed against the hybrid design matrix. This method was inspired by the various corrector techniques available from \texttt{lightkurve} and makes heavy use of its \texttt{RegressionCorrector} class\footnote{\url{https://docs.lightkurve.org/reference/api/lightkurve.correctors.RegressionCorrector.html}}. %It is a preliminary version of a software package being developed for AGN using TESS data, \texttt{Quaver}, scheduled to be released in the near future.

Figure 1 % ~\ref{fig:lc}
shows the light curves of the sources analyzed in this work. The upper left panel shows the light curve of BL Lacertae. \textsl{TESS} observed this source beginning on 12 September 2019 for around 24.6 days in sector (Sec.) 16 of Cycle 2 observations. The gray curves represent the error associated with the fluxes which are shown in blue. The red curve is the running average of the light curve taken with the \texttt{window=20} to highlight the possibly periodic features present in the data  (\texttt{window} defines the number of measurements for which the running average is calculated). The upper right panel shows the light curve of 1RXS J004519.6+212735. This BL Lac object was observed beginning on 8 October 2019 for about 23.7 days in Sec.\ 17 of Cycle 2 observations. Both lower panels belong to 1RXS J111741.0+254858. The Sec.\ 22 observation of Cycle 2 began on 21 February 2021 and continued for about 25.6 days; it is shown on the lower left panel of Fig.\ 1. The lower right panel shows the light curve of the Sec.\ 49 observation of Cycle 4 of 1RXS J111741.0+254858 starting on 28 February 2022 and continuing for about 24.8 days.
Figure~2~%\ref{fig:sample_stars}
shows the light curve of two inactive stars from the same postage stamps as 1RXS J004519.6+212735 and 1RXS J111741.0+254858, reduced using the same pipeline. The reduced light curves are flat, indicating our successful removal of systematics.

\section{Data Analysis}\label{sec:analysis}

In this section, we discuss the timing analysis techniques used in this work to search for any periodicities or quasi-periodicities and to assess their significance in the four light curves shown in Fig.\ 1. %fig.~\ref{fig:lc}.
These methods are the generalized Lomb-Scargle periodogram and weighted wavelet Z-transform analyses.
\subsection{Generalized Lomb-Scargle Periodogram }

The Lomb-Scargle periodogram (LSP) is a powerful Fourier transform frequently used to detect periodicity in unevenly sampled data ~\citep{1976Ap&SS..39..447L, 1982ApJ...263..835S}. It uses likelihood minimization to fit sinusoidal components throughout the data. The Generalized LSP (GLSP) is an advanced version of the `classical' LSP which includes the error of fluxes in the periodogram calculations and fits the data with both sinusoidal and constant components.

In our work, we employ the GLSP routine provided by {\tt pyastronomy}\footnote{ https://github.com/sczesla/PyAstronomy}. It applies the method described in \cite{2009A&A...496..577Z}. In the code, the frequency resolution is governed by the oversampling factor which is needed for  the calculation of the power spectrum and its significance. Here, we took the oversampling factor to be 2.0, which corresponds to the number of frequencies considered to be equal to the number of data points in a single observation. We have confirmed that the result is insensitive to different oversampling factors, by also employing oversampling factors of 1.0 and 10.0. We found that the values of the  frequencies of the peaks of the signals exceeding 3$\sigma$ significance vary by no more than 3\% for that range of oversampling factors and the value of the powers at those peaks are within just a few percent of each other.

\subsection{Weighted wavelet Z-transform analysis}
In addition to detecting any possible periodicity in a signal, it is also imperative to assess its persistence
throughout the observation. The wavelet method decomposes the signal into both frequency and time domains simultaneously. Instead of the sinusoidal components, this method uses wavelet functions to fit the observation. In this work, we employed the weighted wavelet Z-transform (WWZ) method which can be used effectively for unevenly sampled and sparse data ~\citep{2005NPGeo..12..345W}. It uses the z-statistics for the calculations \citep{1996AJ....112.1709F} and is based on the Morlet wavelet function  which is the function of both frequency and time. In this work, we use the publicly available WWZ\footnote{WWZ software is available at URL: https://www.aavso.org/software-directory.} software that has been used extensively in the literature \citep[see][and references therein]{ 2017ApJ...847....7B, 2017ApJ...849....9Z,2021MNRAS.501.5997T}.

The major assumption in wavelet calculations is the type of wavelet function used. The finite length of the observation can induce the edge effects which could be observed in the density plots. We also note that the frequency and temporal resolutions can have important effects on the calculation of the significance of peaks.

\subsection{Significance Calculations}

\begin{figure*}
\centering
\includegraphics[scale = 0.3]{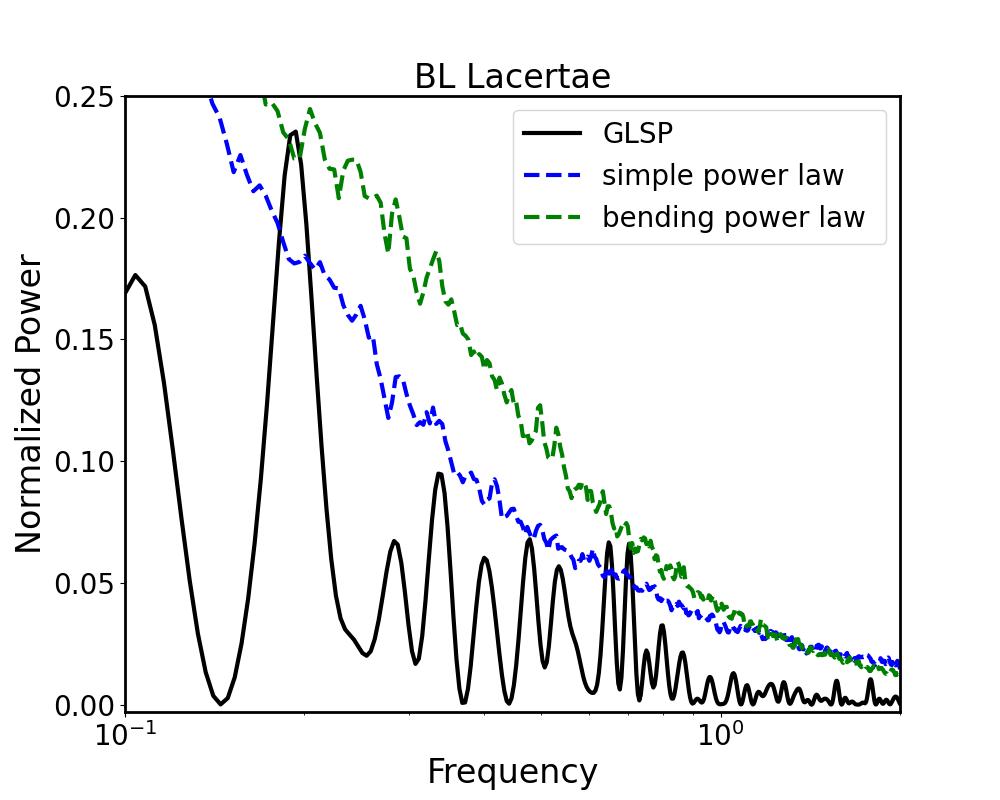}\includegraphics[scale = 0.3]{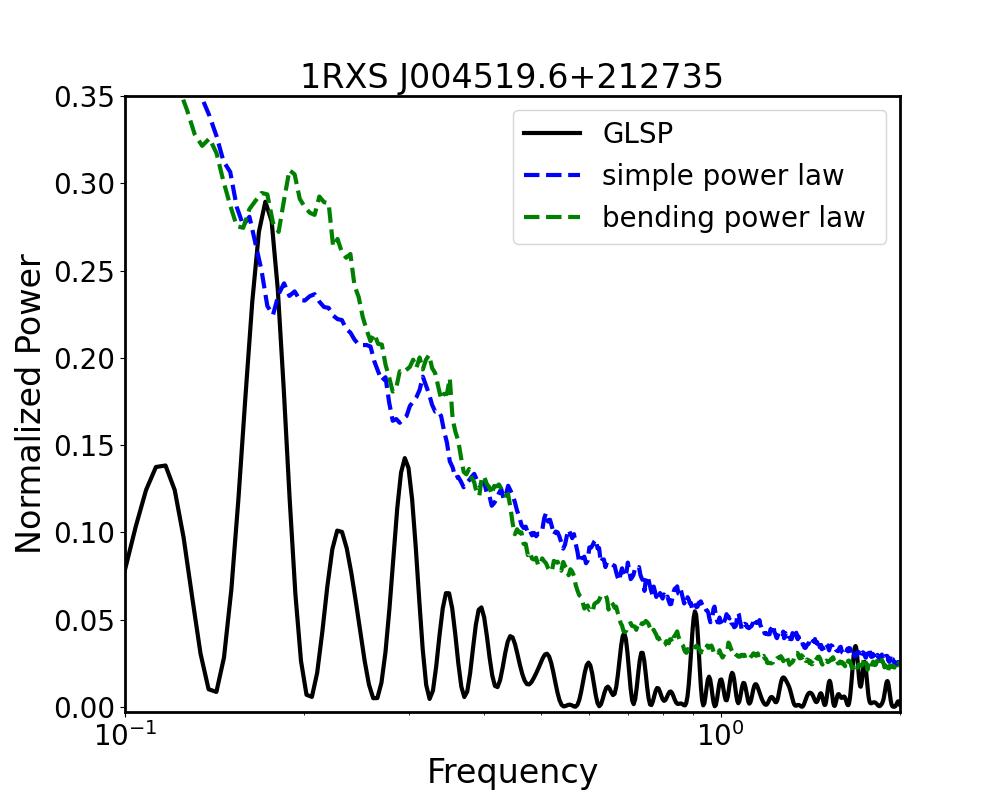}\\
\includegraphics[scale = 0.3]{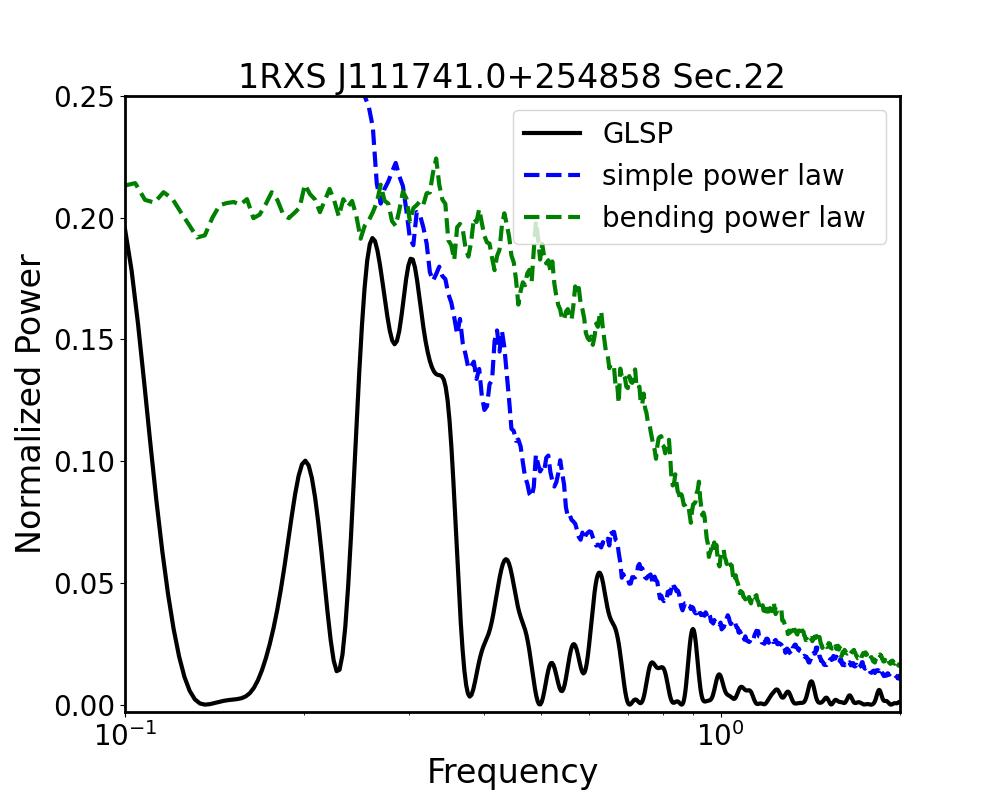}\includegraphics[scale = 0.3]{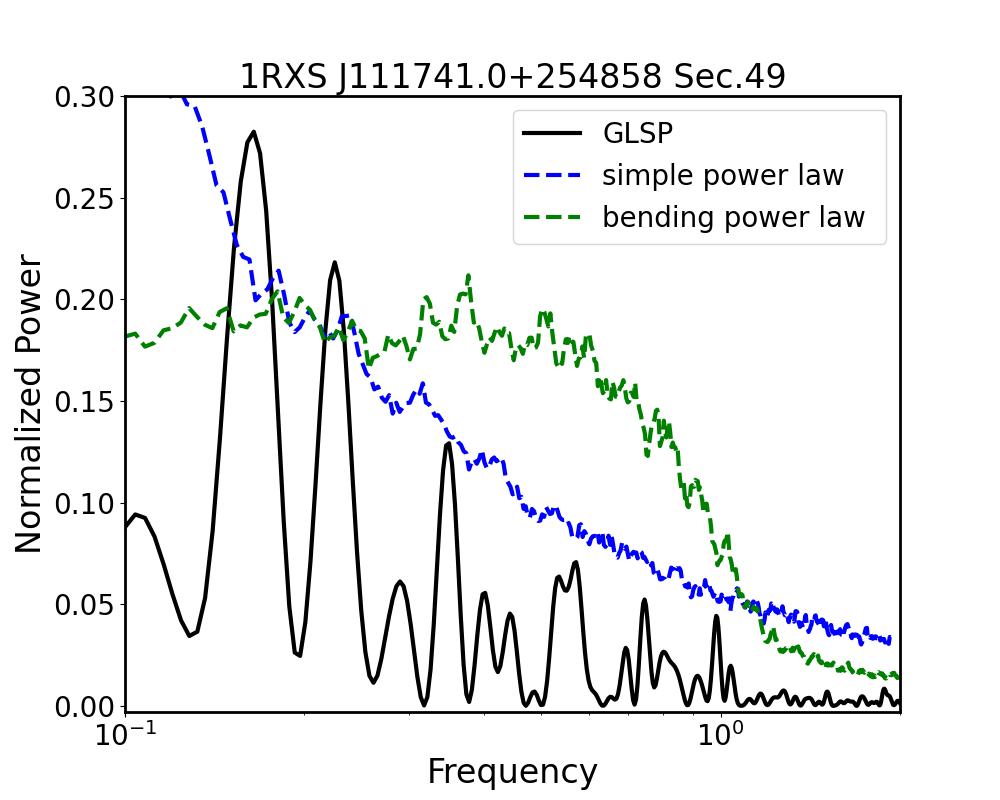}\\
\caption{Generalized Lomb-Scargle Periodograms for the \textsl{TESS} light curves analyzed in the work. The black histogram is the power spectral density, the blue curve shows 3$\sigma$ significance for simple power law and green curves shows the same confidence for bending power law. {\it Upper panel}: BL Lacertae (left) and 1RXS J004519.6+212735 (right). {\it Lower panel}:  Cycle 2 (left) and Cycle 4 (right) observations of 1RXS J111741.0+254858.}
\end{figure*}\label{fig:glsp}

%\begin{figure}
%\centering
%\includegraphics[scale = 0.45]{glsp_16.eps}\includegraphics[scale = 0.45]{glsp_17a.eps}\\
%\includegraphics[scale = 0.45]{glsp_22.eps}\includegraphics[scale = 0.45]{glsp_49.eps}\\
%\caption{Generalized Lomb-Scargle Periodograms for the \textsl{TESS} light curves analyzed in the work. The black histogram is the power spectral density, \textcolor{red}{the red curve is the best-fit bending power law and} the blue curve shows 3$\sigma$ significance. {\it Upper panel}: BL Lacertae (left) and 1RXS J004519.6+212735 (right). {\it Lower panel}:  cycle 2 (left) and cycle 4 (right) observations of 1RXS J111741.0+254858.}
%\end{figure}\label{fig:glsp}

%\begin{figure}
%\centering
%\includegraphics[scale = 0.45]{glsp_16.png}\includegraphics[scale = 0.45]{glsp_17.png}\\
%\includegraphics[scale = 0.45]{glsp_22.png}\includegraphics[scale = 0.45]{glsp_49.png}\\
%\caption{Generalized Lomb-Scargle Periodograms for the \textsl{TESS} light curves analyzed in the work. The blue histogram is the power spectral density, \textcolor{red}{the red curve is the best-fit bending power law and} the green curve shows 3$\sigma$ significance. {\it Upper panel}: BL Lacertae (left) and 1RXS J004519.6+212735 (right). {\it Lower panel}:  cycle 2 (left) and cycle 4 (right) observations of 1RXS J111741.0+254858.}
%\end{figure}\label{fig:glsp}

    It appears that the stochastic processes occurring in both turbulent jets and accretion disks  usually manifest themselves in the light curves of blazars as is seen in  observations taken throughout the electromagnetic spectrum \citep[e.g.][]{2014MNRAS.445..437M, 2018A&A...620A.185N, 2020ApJ...891..120B}. A blazar Power Spectrum Density (PSD), $P(f)$, as a function of frequency $f$, often follows a simple power-law of the form $f^{\alpha}$ where $\alpha$ is the spectral index for a significant range of frequencies. Any QPO, on the other hand, is believed to arise from coherent disk or jet processes. So, a strong QPO feature should project itself in the blazar PSD above the power-law. It is imperative to estimate the significance of the peak relative to  the red noise of the PSD. As $\alpha$ usually has a negative value, meaning that the spectrum at lower frequencies fluctuates with larger amplitudes, it is possible to misinterpret these fluctuations as quasi-periodic signals. In addition, there may be some fluctuations introduced in the signal by  processes arising from the observational constraints which could be misinterpreted as periodic peaks. Therefore, it is very important to estimate the significance of these peaks.

To perform significance calculations, we first simulate light curves having the same properties as the observation (mean, variance, duration) and power spectrum. Using the standard discreet Fourier Transform (DFT) for the unevenly sampled data as used in this work would be problematic as it results in amplitude fluctuation and unreasonable frequency shifting \citep[e.g.][]{2017ApJ...837...45F}. Although the \textsl{TESS} data used in this work are far more uniformly spaced than most observations, they do contain some gaps, so the observations are not taken at strictly regular intervals. To avoid these issues, we employ the GLSP method. The light curve analyzed in this work is essentially two segments separated by a rather wide $\sim$4-day gap (except for BL Lac, where the gap is only $\sim$1 day). To simulate these light curves realistically, we essentially treat segments individually. The resultant simulated light curves consisted of the two $\sim$10-day segments with $\sim$4-day gap in the middle.  We chose the frequency spacing such that the number of frequencies is equal to the number of data points in the observation. We fit the power spectrum $P(\nu)$ at a given frequency $\nu$ with a bending power-law plus a constant representing the instrumental white noise that dominates at high frequencies
\begin{equation}
   P(\nu) = N \left(1+ \left(\frac{\nu}{\nu_b} \right)^{-\beta}\right)^{-1.0} + C,
\end{equation}
where $N$ is the normalization, $\nu_b$ is the bending frequency, $\beta$ is the power-spectrum index, and $C$ is the constant representing the instrumental white noise.
We used the {\tt curve fit} routine in {\tt scipy} package of {\tt python} to estimate the parameters of the red-noise spectrum. We also fit the power spectrum with simple power law defined as
 \begin{equation}
   P(\nu) = N \left(\frac{\nu}{\nu_0}\right)^{\alpha},
\end{equation}
where $\nu_0$ is the constant. Although a simple power-law and bending power-law are chosen to model the `red-noise' in the spectra, we note that it is possible to explain the stochastic process using different
mathematical models; e.g., short memory ARMA models are capable of producing periodic patterns for AR(p) with p$>$2 \citep[see][and references therein]{2002CG.....28..421S}.

Using the parameters of each power spectrum obtained using GLSP and the other properties of the source, we simulate 10000 light curves employing the {\tt stingray}~\citep{2019ApJ...881...39H} timing analysis package\footnote{https://github.com/StingraySoftware} which implements the algorithm described in \cite{{1995A&A...300..707T}}. We calculate GLSP for the simulated light curves. The significance is estimated from the PSD distribution at each frequency.
%\sout{At each frequency, we take the average of power from all light curves and use it as a baseline to estimate the significance following \cite{2005A&A...431..391V}. All significances calculated in this work are calculated globally, i.e., they include the number of frequencies used in the calculation of periodogram \citep[for details, see][] {2021MNRAS.501.5997T}. Including the significance solely from the distribution of power at each frequency but not including the number of frequencies would overestimate the significance. }

We also perform Monte Carlo Markov Chain (MCMC) simulations to estimate errors on the best-fit parameters of the best-fit bending power law.
%Fig.~4 shows the corner plots of the posterior distribution of parameters associated with the  bending power laws sampled from 100,000 iterations and 500 walkers.
Table 1 and Table 2  show the best-fit parameters and the associated errors for the bending power-law model and the simple power law, respectively.

To estimate the significance of the signal in the WWZ method, we calculate the time-averaged WWZ, which is essentially the WWZ power marginalized over time. This time-averaged WWZ is a periodogram that follows the $\chi^2$ distribution with 2 degrees of freedom in the limit of even sampling \citep{1996AJ....112.1709F}. We calculate time-averaged WWZ for 10,000 simulated light curves. The significance, as in the case of GLSP, is derived from the distribution of the simulated time-averaged WWZ at each frequency.

\begin{figure*}
\centering
\includegraphics[angle=90,scale = 0.35]{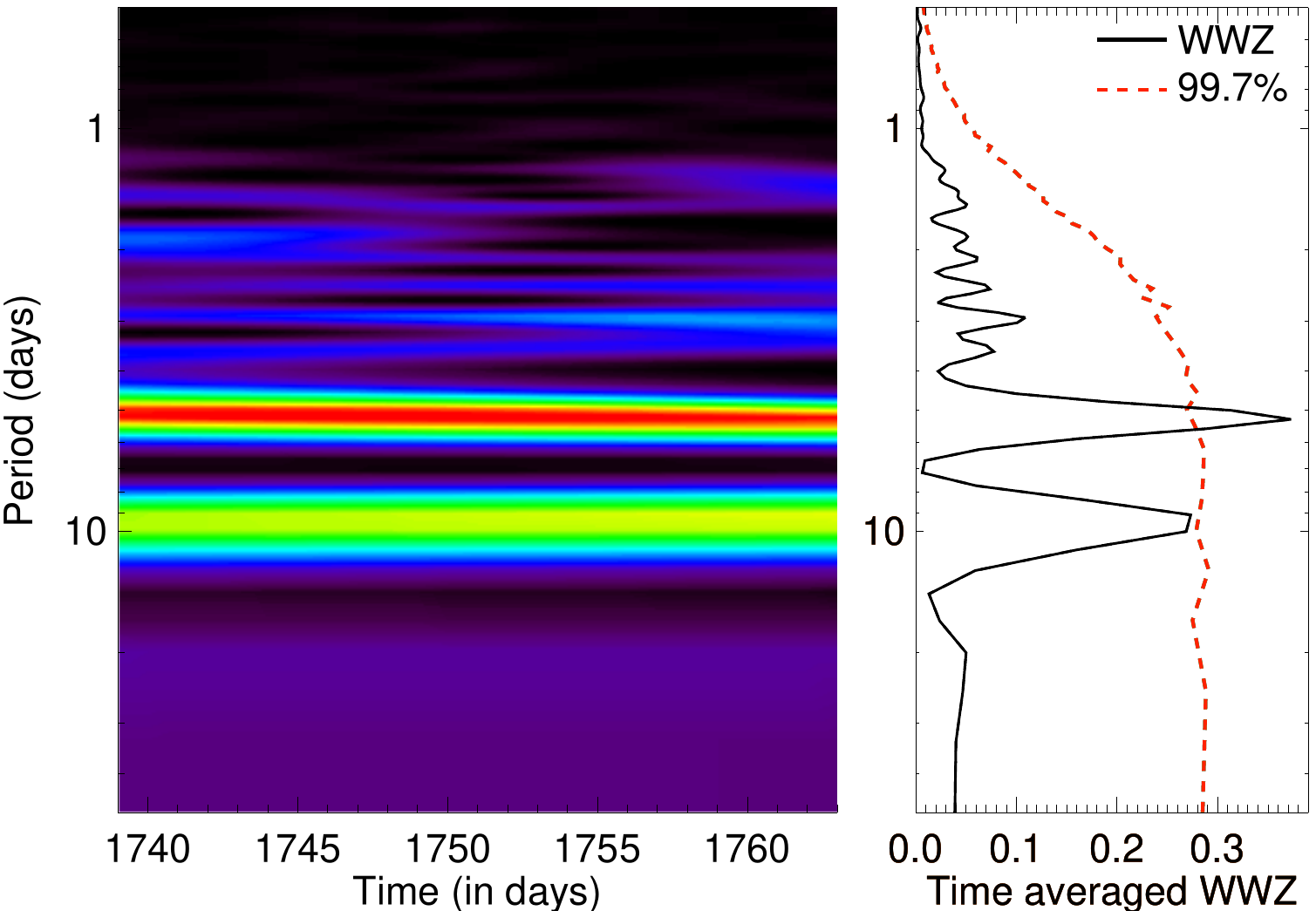}\includegraphics[angle=90,scale = 0.35]{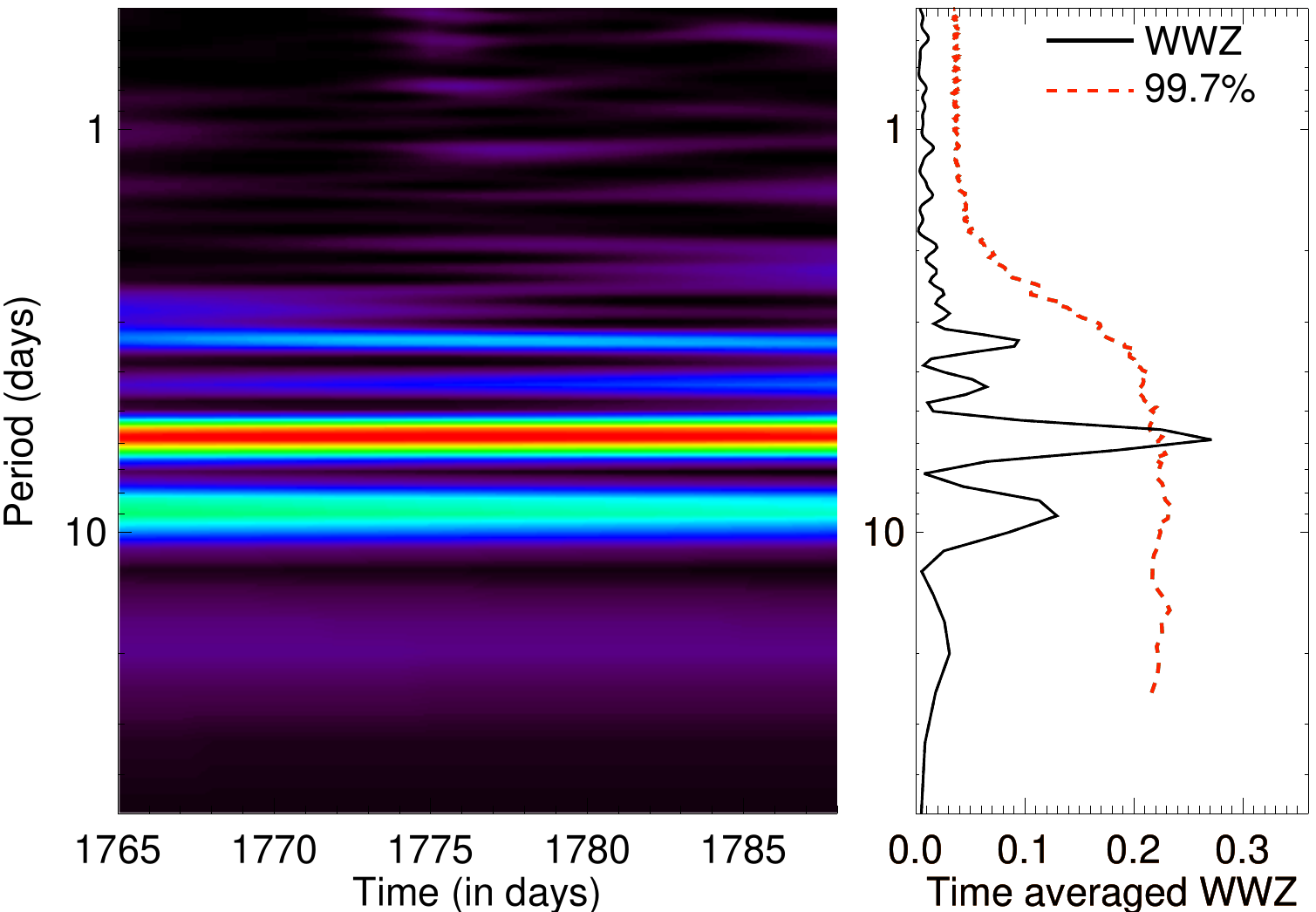}\\
%\vspace{-0.7cm}
\includegraphics[angle=90,scale = 0.35]{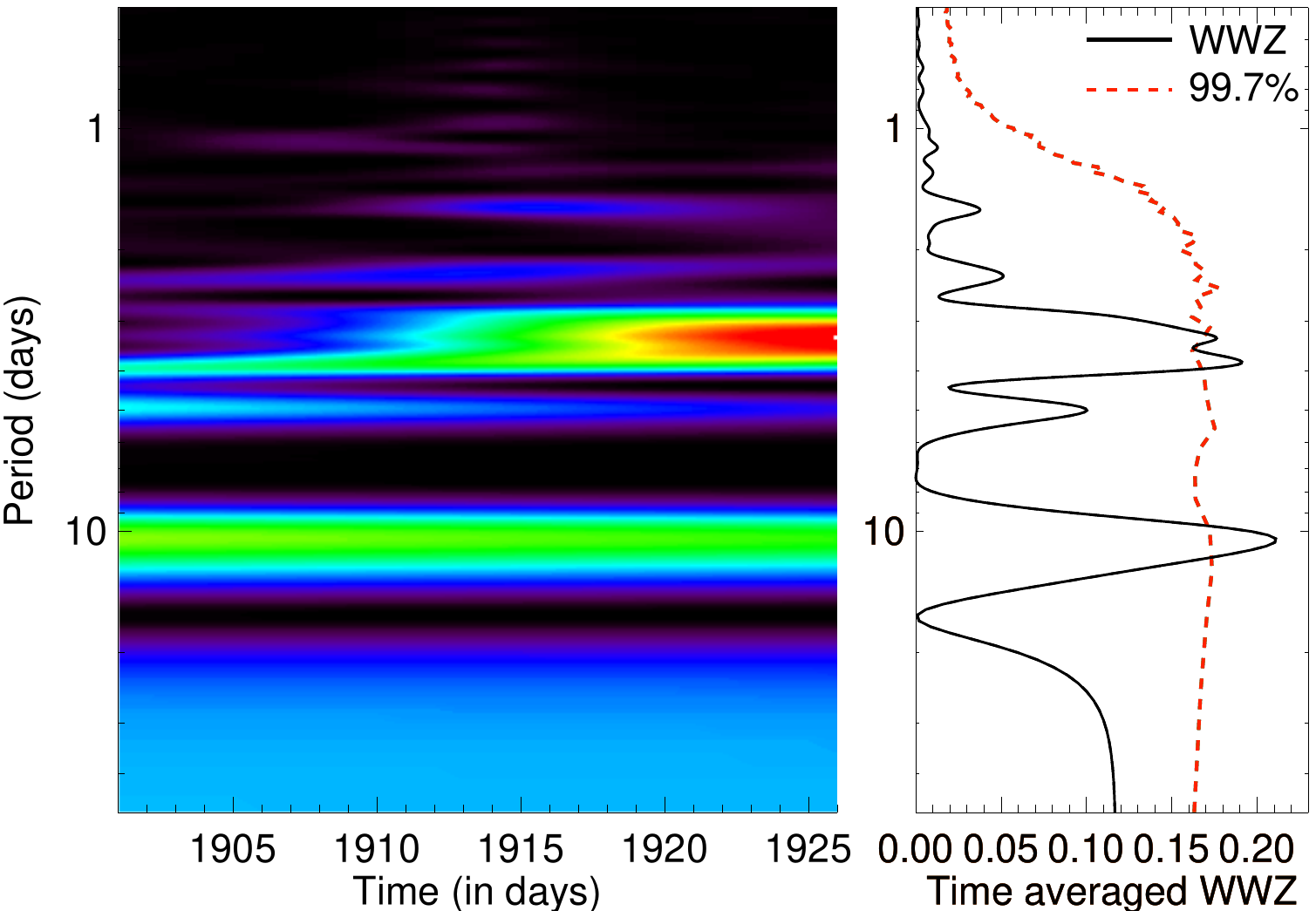}\includegraphics[angle=90,scale = 0.35]{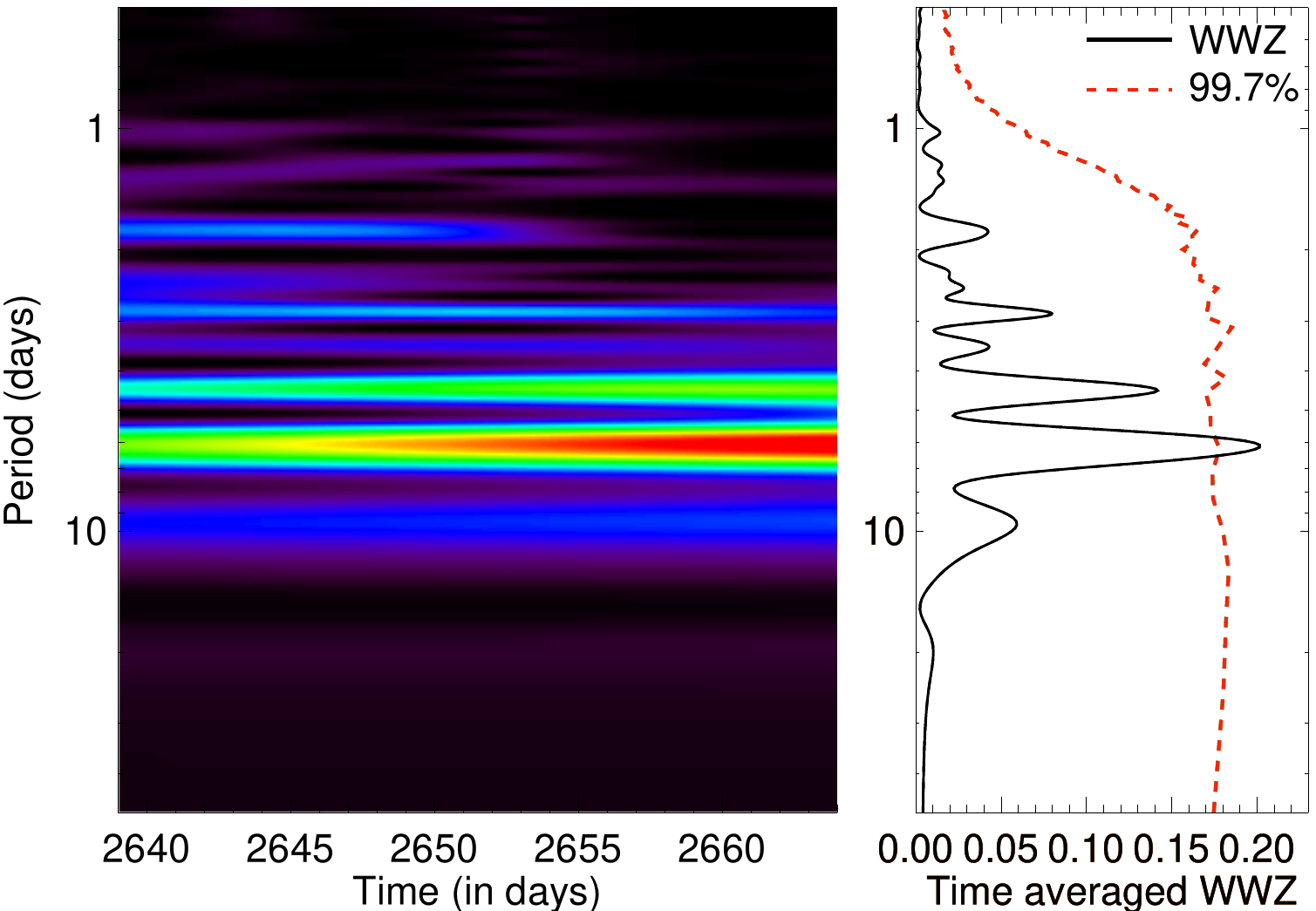}\\
\caption{Weighted wavelet Z-transform (WWZ) analysis results for the light curves analyzed in the work. In each plot, the left sub-panel corresponds to the color-color diagram of wavelet power (red  denoting the maximum power and decreasing towards violet and black) plotted against time and frequency. The right sub-panel shows the time-averaged WWZ plotted against frequency in black; the red dotted lines represent 3$\sigma$ significances. {\it Upper panel}: BL Lac (left) and 1RXS J004519.6+212735 (right). {\it Lower panel}: 1RXS J111741.0+254858 in Cycle 2 (left) and Cycle 4 (right).}
\end{figure*}\label{fig:wwz}

\section{Results}\label{sec:res}
Fig.\ 1 %~\ref{fig:lc}
 shows the light curves we obtained from the \textsl{TESS} observations
 for BL Lac (upper left panel), 1RXS J004519.6+212735 (upper right panel), and 1RXS J111741.0+254858 (lower panels). Visually, there seem to be possible quasi-periodic components that were searched for and confirmed by the GLSP and WWZ methods.  %generalized Lomb-Scargle periodogram (GLSP), REDFIT analysis, and weighted wavelet Z-transform (WWZ) analysis methods.
These methods were chosen as they provide distinctive approaches to the time series data.  Explicitly, GLSP gives the Fourier transform of the data, probing the frequency domain while the wavelet analysis decomposes the data into frequency and time simultaneously. Only if a peak shows at least  3~$\sigma$ significance in both the GLSP and time-averaged WWZ methods do we consider it to be a solid QPO signal.

%figure~\ref{fig:glsp}
Fig.\ 3 shows the normalized power (black curve) of the GLSP plotted against frequency for BL Lac. The blue curves denote the 3$\sigma$ (99.73\%) significance obtained by simulations assuming simple power law as the underlying model. The green curve corresponds to 3$\sigma$ significance obtained by simulations assuming bending power law. The most prominent peak, with a significance of more than 3~$\sigma$, occurs at the period of 4.94 days. The peak around 1.40 days is also prominent ($>$ 3 $\sigma$ significance) for both bending and simple power law.

Note that quoting a period of more than 6 days would correspond to having fewer than 4 cycles for the timescales of our observations and therefore are not suitable for claiming a QPO. This limitation is supported by early work which showed that it is possible to apparently see 3 cycles in many light curves based purely on flicker noise \citep{1978ComAp...7..103P}. \citet{2016MNRAS.461.3145V} found that greater numbers of cycles that follow a periodic pattern more  strongly support the presence of periodicity. Even a pure sinusoidal component could not be distinguished from a simple stochastic process if there are only $\sim$~2 cycles. However, the presence of $\sim$~5 cycles could be easily distinguished from a stochastic process.

%We next consider the results of the auto-regressive analysis. The upper left plot of Fig.\ 4 %Fig.~\ref{fig:redfit}
%provides the REDFIT results for the BL Lac light curve binned at 3 hours. The black, red, blue, and magenta curves represent the bias-corrected spectrum, theoretical AR(1) spectrum, 95\% significance, and 99\% significance, respectively. \textcolor{red}{ Peaks with a significance of more than 99\% are detected at periods of 1.25, 1.42, and 1.57 days. The peaks at 2.96 days and 5.28 days have  significances} of more than 95\%. The remaining peaks have significance no greater than 95\% confidence. \\
%\sout{In all four cases,***BUT THIS DOESN'T SEEM TO BE THE CASE FOR 1RXS J111741 IN SECTOR 22*** the strongest acceptable frequency is at least 3 $\sigma$ significant in the GLSP analysis assuming either a simple power law or a bending power law as the underlying model.}
As the time-averaged WWZ is also effectively a periodogram that contains similar information as the GLSP, showing only one underlying model in the WWZ plots is sufficient. We note that the claimed frequencies in the Fourier space lie in the "red-noise" regime where the PSD can be adequately described by both simple power-law and bending power-law models. So we show the WWZ significance plots only for  bending power laws.

The upper left panel of Fig.\ 4 %\ref{fig:wwz}
shows the result of our WWZ analysis for BL Lac.
The left plot shows the color-color diagram of WWZ power plotted against time and frequency. %(red being the most concentrated power and decreasing towards violet). ***REDUNDANT WITH CAPTION***
The maximum power is concentrated around the period of 5.26 days, and it persists throughout the observation. The second most prominent concentration,  which also persists throughout the observation, is around 2.94 days. There are other weaker features that seem to be present in different parts of the observation.
For instance, the feature around 1.54 days becomes stronger in the later half of the observation whereas the feature around 2.0 days is only visible in the first half. The black curve in the  right sub-panel of Fig.\ 5 for BL Lac %Fig.\ref{fig:wwz}
shows the time-averaged WWZ and the red dotted curve denotes the 3$\sigma$ (99.73\%) confidence curve. The feature at 5.26 days, which also appears in the WWZ  plot, is found to be above 3$\sigma$. The signal around 5 days is found to be significant in both methods. However, as it corresponds to fewer than 5 cycles during the 24.6-day TESS observations, it cannot be considered to be very convincing.

The upper right plot of Fig.\ 3 %fig.~\ref{fig:glsp}
 shows the GLSP for Cycle 2 data of 1RXS J004519.6+212735. The peaks at the period of 5.8 days are marginally consistent with 3~$\sigma$ significance. The color-color diagram of the WWZ analysis of this observation (upper right panel of Fig.\ 4) shows  that the peak around 5.88 days is persistent throughout the observation and is the only peak having significance more than 3$\sigma$ in the time-averaged WWZ plot on the right-hand side of the panel. Although the peak at 5.9 days is nominally the most significant measurement  from both methods, it corresponds to only 4.0 cycles during the length of the observations, so cannot be fully trusted.
 
 %The peak around 3.4 days is the second most  significant \textcolor{red}{(but less than at 3~$\sigma$) but} it corresponds to 7 cycles.
\begin{table*}
 \centering
 \caption{Best fit parameters and associated errors for the bending power-law sampled from MCMC realizations. }
\begin{tabular}{|c|c|c|c|c|c|}
\hline
Source \& Sector & Segment & log $N$ & $\beta$ & $\nu_b$ & $C$\\\hline
BL Lacertae Sec.\ 16 &-& $-0.95^{+0.09}_{-0.09}$& $-1.80^{+0.15}_{-0.16}$ & $0.27^{+0.02}_{-0.02}$ & $0.0012^{+0.0001}_{-0.0001}$ \\ \hline
{1RXS J004519.6+212735 Sec.\ 17 } & 1 & $-0.33^{+0.05}_{-0.05}$& $-3.21^{+0.28}_{-0.30}$ & $0.24^{+0.02}_{-0.02}$ & $0.01^{+0.001}_{-0.001}$ \\
 & 2 & $-0.85^{+0.08}_{-0.07}$& $-3.33^{+0.29}_{-0.29}$ & $0.32^{+0.03}_{-0.03}$ & $0.009^{+0.0008}_{-0.0008}$ \\ \hline
{1RXS J111741.0+254858 Sec.\ 22} & 1 & $-1.44^{+0.13}_{-0.13}$& $-3.39^{+0.29}_{-0.29}$ & $0.65^{+0.06}_{-0.05}$ & $0.009^{+0.001}_{-0.001}$ \\
 & 2 & $-1.68^{+0.14}_{-0.14}$& $-6.12^{+0.56}_{-0.53}$ & $0.81^{+0.07}_{-0.07}$ & $0.002^{+0.001}_{-0.001}$ \\ \hline
1RXS J111741.0+254858 Sec.\ 49 & 1 & $-1.15^{+0.10}_{-0.11}$& $-5.61^{+0.51}_{-0.49}$ & $0.81^{+0.07}_{-0.07}$ & $0.003^{+0.0003}_{-0.0003}$ \\
 & 2 & $-1.62^{+0.14}_{-0.14}$& $-8.92^{+0.80}_{-0.81}$ & $0.96^{+0.09}_{-0.08}$ & $0.002^{+0.0002}_{-0.0002}$ \\ \hline
\end{tabular}
\end{table*}

\begin{table*}
 \centering
 \caption{Best fit parameters and associated errors for the simple power-law sampled from MCMC realizations. }
\begin{tabular}{|c|c|c|c|c|}
\hline
Source \& Sector & Segment & log $N$ & $\alpha$ & $\nu_0$ \\\hline
BL Lacertae Sec.\ 16 &-& $-1.82^{+0.16}_{-0.16}$& $-1.11^{+0.18}_{-0.19}$ & $0.65^{+0.06}_{-0.06}$  \\ \hline
{1RXS J004519.6+212735 Sec.\ 17} & 1 & $-1.41^{+0.13}_{-0.13}$& $-1.21^{+0.11}_{-0.11}$ & $0.69^{+0.06}_{-0.06}$  \\
 & 2 & $-1.40^{+0.12}_{-0.13}$& $-0.85^{+0.08}_{-0.07}$ & $0.68^{+0.06}_{-0.06}$ \\ \hline
{1RXS J111741.0+254858 Sec.\ 22} & 1 & $-1.22^{+0.11}_{-0.10}$& $-1.24^{+0.11}_{-0.10}$ & $0.38^{+0.03}_{-0.03}$  \\
 & 2 & $0.62^{+0.06}_{-0.05}$& $-2.16^{+0.19}_{-0.19}$ & $0.06^{+0.01}_{-0.01}$ \\ \hline

{1RXS J111741.0+254858 Sec.\ 49} & 1 & $-1.32^{+0.11}_{-0.12}$& $-0.62^{+0.05}_{-0.06}$ & $0.40^{+0.03}_{-0.04}$  \\
 & 2 & $0.04^{+0.01}_{-0.01}$& $-1.14^{+0.11}_{-0.12}$ & $0.06^{+0.01}_{-0.01}$ \\ \hline
\end{tabular}
\end{table*}
We now discuss the results obtained for  1RXS J111741.0+254858. There are two observations of this source, one from sector 22 of Cycle~2 and one from sector 49 of Cycle~4. The lower left panel of Fig.\ 3 %fig.~\ref{fig:glsp}
shows  the %evolution of     GLSP with time
 GLSP for Cycle 2 data. The most prominent period (at less than 6 days) is found at 3.68 days. The feature around 3.68 days has a significance of $\approx$ 3$\sigma$ and has a double-horned structure which is confirmed by the other method. Note that we report the period of the `horn' of this broad peak that has the higher significance. The double horned structure is also seen in the color-color diagram of the wavelet of this Cycle~2 observation (lower left panel of Fig.\ 4 %fig.~\ref{fig:wwz})
 at 3.78 days (higher peak). Interestingly, only one of the peaks is present in the first half of the observation but as the observation evolves with time, the second one appears. So, the behavior of this double-horned peak could only be distinguished when the observation is decomposed into time and frequency domains simultaneously which is achieved by using WWZ. The feature at the period of 5 days is relatively stronger at the beginning of the observation but gets weaker with time and becomes negligible by the end of the observation. All these methods  show the most significance for a possible QPO at around 3.8 days.

The GLSP for the Cycle~4 data of 1RXSJ111741.0+254858 is plotted in the lower right panel of %figure~\ref{fig:glsp}.
Fig.\ 3.  The most prominent peak is found at the period of 6.20 days (although that is over 25\% of the length of the observations, so not convincing) followed by the peaks at 4.38 and 2.86 days. The perseverance of these peaks can be assessed by the wavelet analysis which is plotted in the lower right panel of Fig.\ 4 %fig.~\ref{fig:wwz}
 for Cycle~4 data. The feature at 6.25 days, which is more than 3$\sigma$ significant, seems to be strongly persistent throughout the observation, though it corresponds to too few cycles to be strongly supported. The features at 4.64 days continue to present strongly throughout the observation which is also the case for the somewhat weaker signal at 2.87 days. So, the most promising possible QPOs indicated by both GLSP and WWZ methods are around 6.2, 4.5, and 2.9 days, and the latter would correspond to over 9 cycles.

\subsection{Caveats}
\begin{figure*}
\centering
\includegraphics[scale = 0.25]{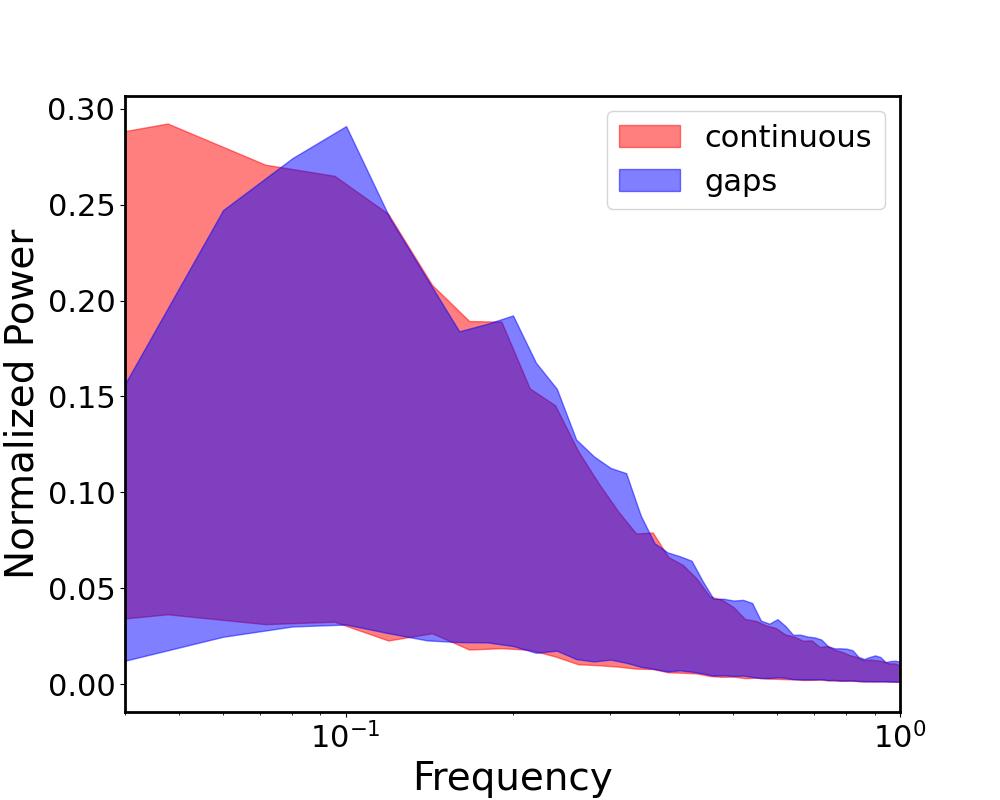}\includegraphics[scale = 0.25]{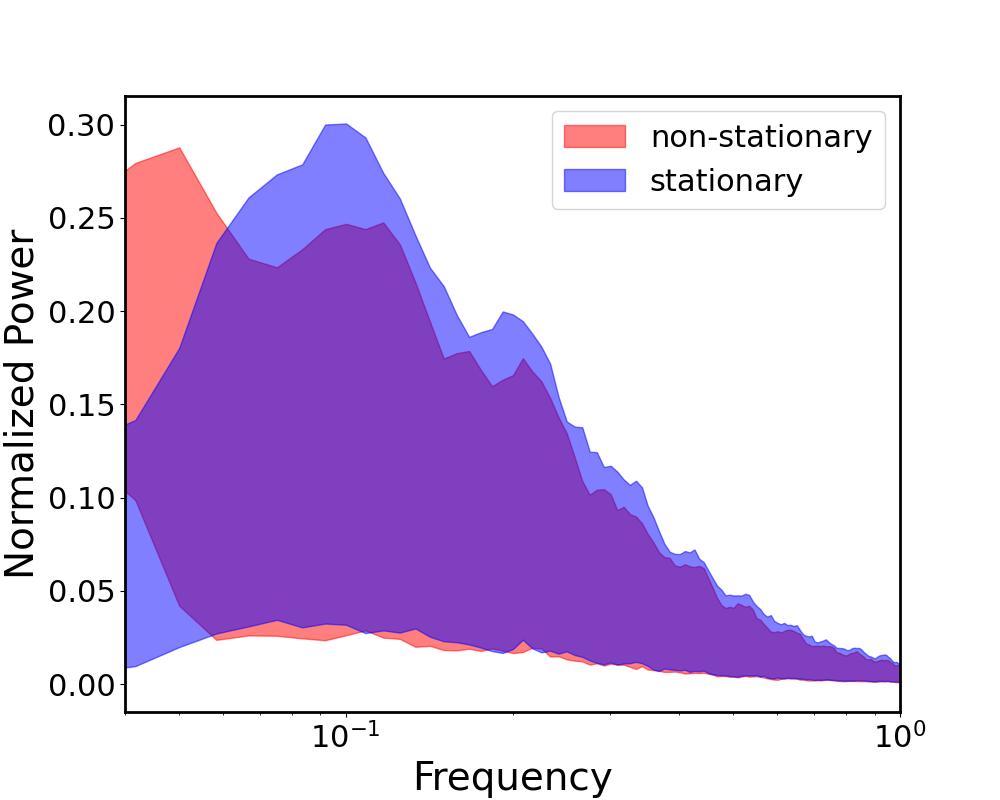}\includegraphics[scale = 0.25]{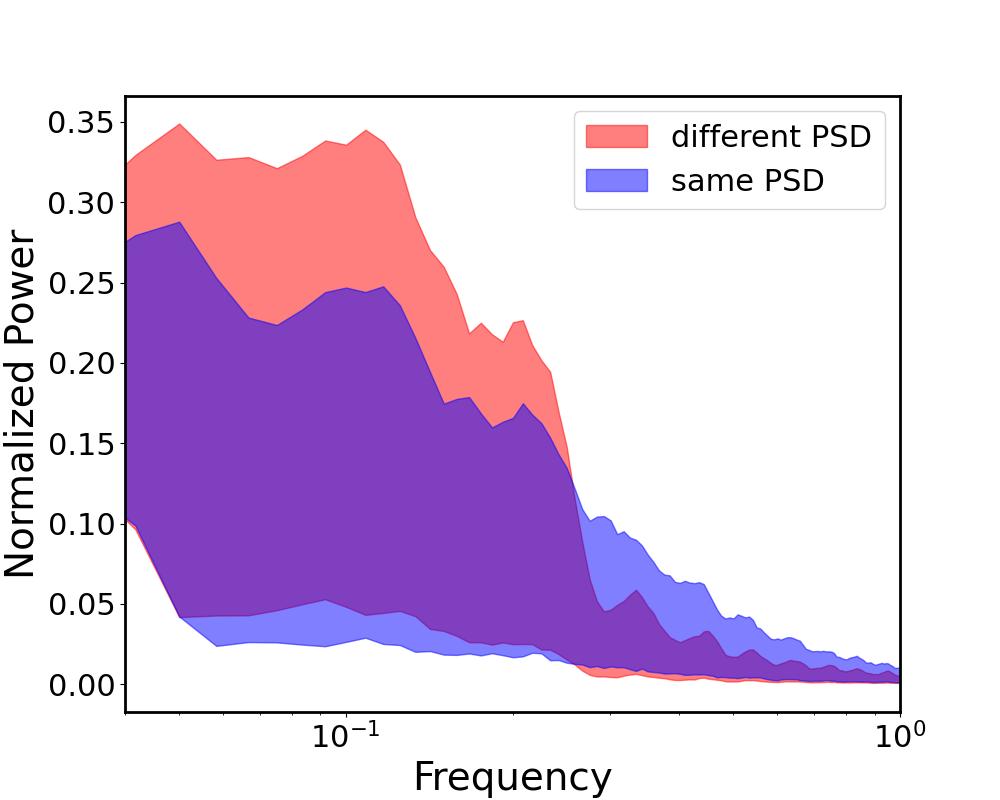}
\caption{Simulated results demonstrating the effects of caveats for a bending power law as the underlying model. {\it Left}: 1 $\sigma$ confidence region for the periodogram obtained by simulating the 20-day light curves; with a gap (blue) and without a gap (red) of 4 days. {\it Middle}: 1 $\sigma$ confidence region for the power spectrum obtained by simulating stationary (blue) and non-stationary (red) light curves. {\it Right}: 1 $\sigma$ confidence region for the periodogram obtained by simulating light curves having the same (blue) and different (red) PSD index. For all of these simulations, we used the inputs from the Cycle 2 observation of 1RXS J111741.0+254858.}
\end{figure*}\label{fig:glsp}

\begin{figure*}
\centering
\includegraphics[scale = 0.25]{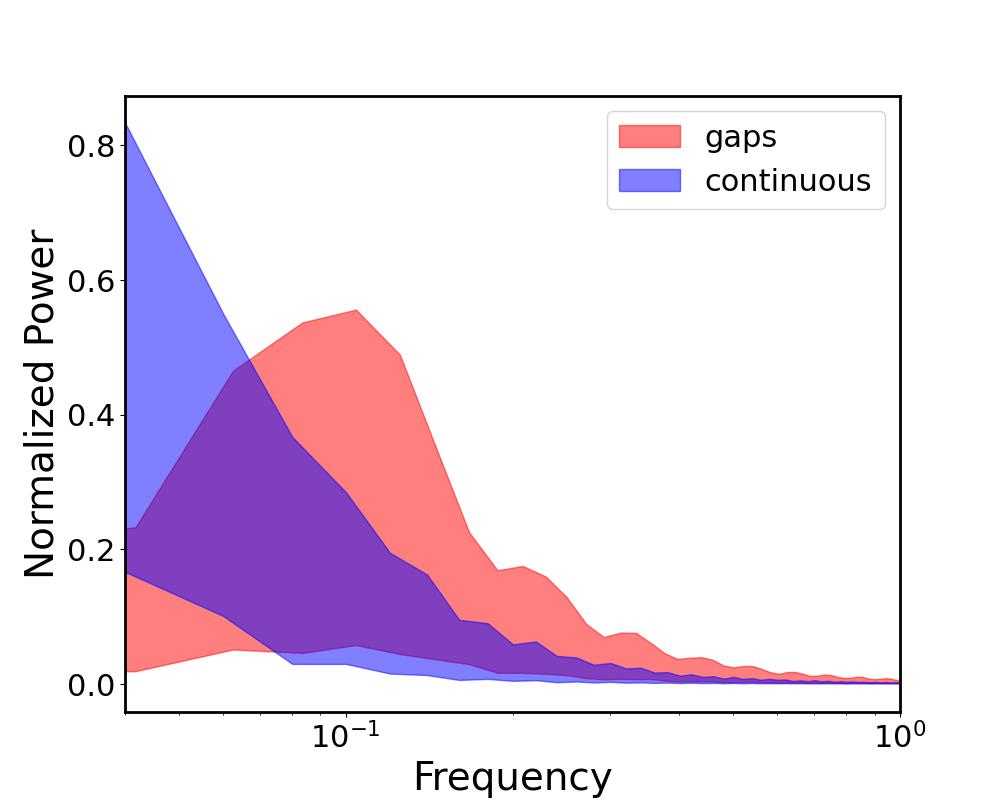}\includegraphics[scale = 0.25]{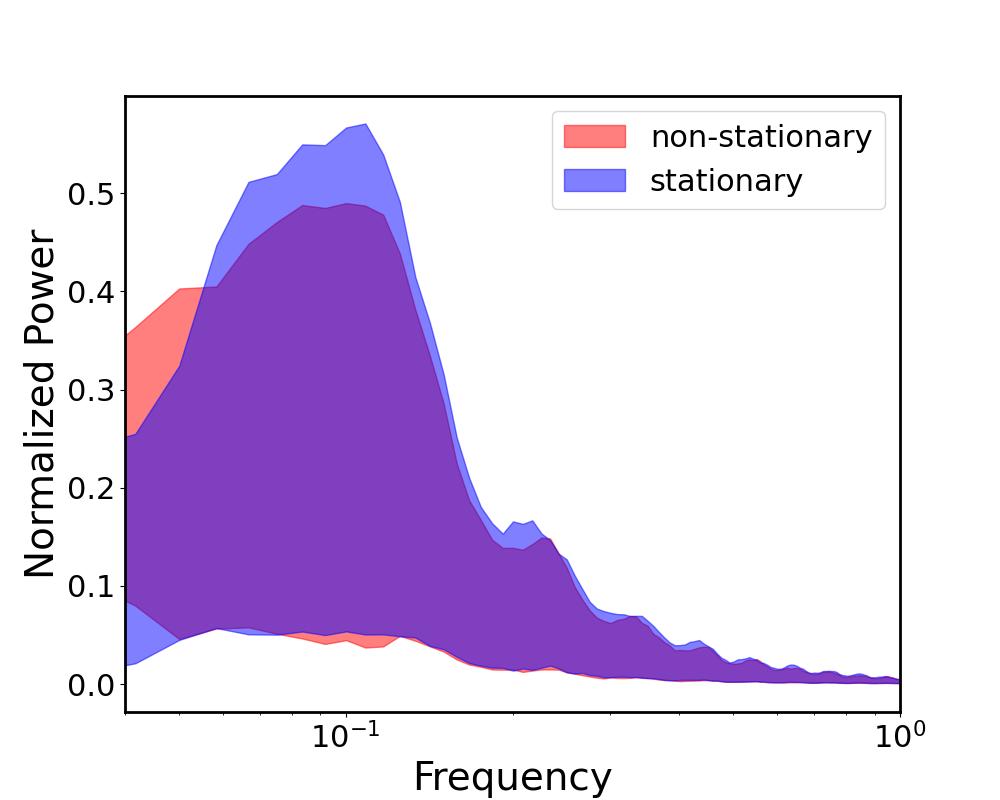}\includegraphics[scale = 0.25]{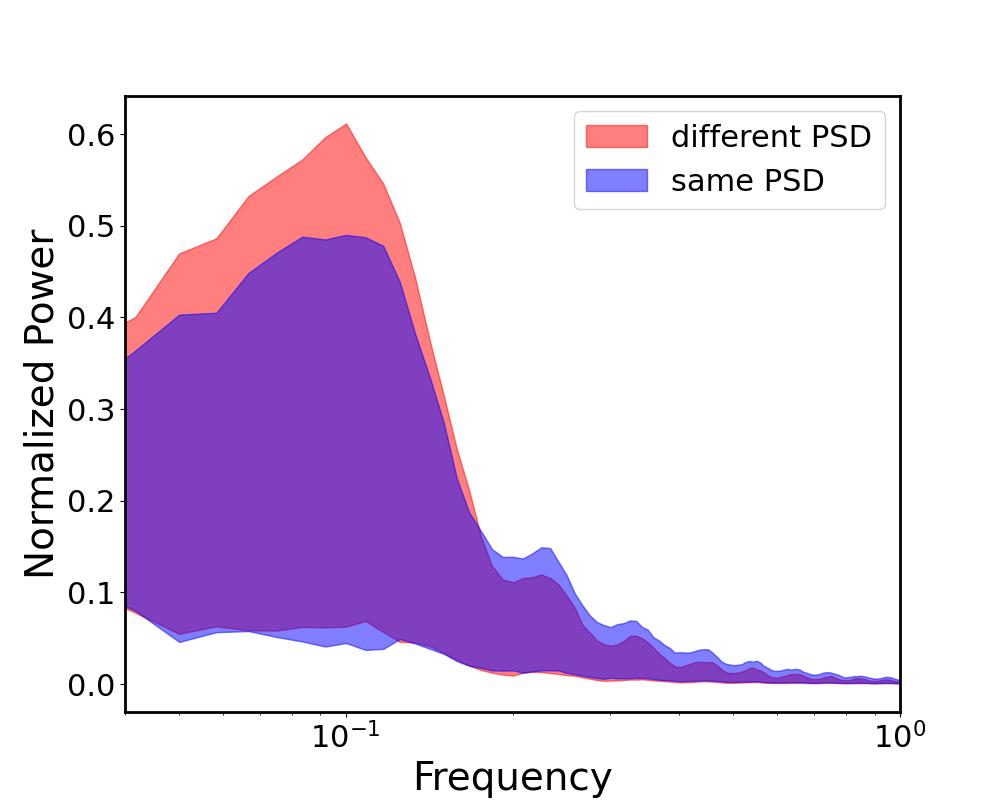}
\caption{ Same as Fig.~5 but for a simple power law as the underlying model. }
\end{figure*}\label{fig:glsp}

One of the main challenges encountered while analyzing \textsl{TESS} light curves is the presence of gaps between light curves which are not negligible in comparison to the duration of the light curves. For instance, a gap of 4 days is present in the 25-day light curve of Cycle 2 of 1RXS J111741.0+254858. To assess the effect of gaps on the power spectrum, we have simulated 1000 light curves each, with and without gaps, and with initial conditions similar to that of Cycle 2 observation of 1RXS J111741.0+254858. The left panel of Fig.\ 5 shows the $1\sigma$ confidence bound for the periodogram for the 20-day light curves both without gaps and with a gap of 4 days, assuming a bending power law as the underlying model. We could deduce from the plot that the periodograms in both cases are consistent with each other at higher frequencies. However, as we go towards lower frequencies ($f~\approx~0.1$d$^{-1}$), the periodogram distributions no longer closely agree with each other. The gap between the light curve segments can affect its power spectrum at frequencies lower than 0.1 d$^{-1}$. So, we restrict our analysis to frequencies above 0.1 d$^{-1}$ to avoid this issue. We note that the effect of a gap is more pronounced in the case of simple power law model for the PSD (left panel of Fig.\ 6). The confidence intervals for the light curves with gaps are more stringent as compared to continuous light curves at low frequencies ($f~\approx~0.07$d$^{-1}$) whereas at higher frequencies, the confidence bounds for continuous light curves are less stringent.

A second issue is the possibility of non-stationarity of the data which affects the power spectrum and the resultant underlying model. To address this, we performed wavelet analysis to assess not only the significance of the signal but also its perseverance. %We can also deduce the duration over which the signal was present during the observation. B
So we have simulated 25-day light curves with a  4-day gap in-between its halves for two cases: stationary, with the same statistical properties throughout the light curve; and non-stationary, where the statistical properties are different for the data before and after the gap. For these simulations, we took the same light curves as before (Cycle 2 observation of 1RXS J111741.0+254858). The middle panels of Figs.\ 5 and 6 show the 1$\sigma$ error bound for stationary and non-stationary light curves with gaps for simple power law and bending power law respectively. In both cases, the bounds are quite similar at frequencies above $f\approx 0.2$d$^{-1}$ and only start to significantly diverge below $f\approx 0.1$d$^{-1}$.

%We have also simulated 25-day light curves with a  4-day gap in-between its halves for two cases: stationary, with the same statistical properties throughout the light curve; and non-stationary, where the statistical properties are different for the data before and after the gap. For these simulations, we took the same light curves as before (Cycle 2 observation of 1RXS J111741.0+254858). The middle panels of Figs.\ 5 and 6 show the 1-$\sigma$ error bound for stationary and non-stationary light curves with gaps for simple power law and bending power law respectively. In both cases, the bounds are quite similar at frequencies above $f\approx 0.2$ and only start to significantly diverge below $f\approx 0.1$d$^{-1}$.
The third challenge is the choice of the underlying model. In this work, we employed both simple power-laws and bending power-laws which could provide incomplete descriptions of the data. By limiting the analysis to frequencies above 0.2~d$^{-1}$, we assumed that the power spectrum behaves like the simple power law (or bending power law in some cases) in this regime.

It is, however, also possible that a single model could not explain the behavior throughout the light curve. So, we fit the segments of the light curves separately and then simulate the light curves using the best-fit parameters for each segment. In this case, the confidence intervals with the same spectral index in the underlying bending power law (right panel of Fig.\ 5) are somewhat less tightly bound as compared to the different spectral index cases at higher frequencies; however, at lower frequencies, the confidence intervals are actually tighter for the same case. In the case of simple power laws (right panel of Fig.\ 6), the error bounds agree with each other quite closely throughout, though the different power-law case is somewhat broader. It can be concluded from these simulations that if we use the same spectral index throughout the observation, the confidence bound is underestimated as compared to the case where different spectral indices are used for different segments of light curves. So, it is preferable to simulate the segments within light curves separately if the gap is large in comparison to the resolution of the data.

The timescales found in this work are in the range of 3--5 days, corresponding to frequencies of 0.20--0.34 d$^{-1}$ so the caveats discussed in this section should have little impact.

\subsection{Segment-wise analysis}
\begin{figure*}
\centering
\includegraphics[scale = 0.25]{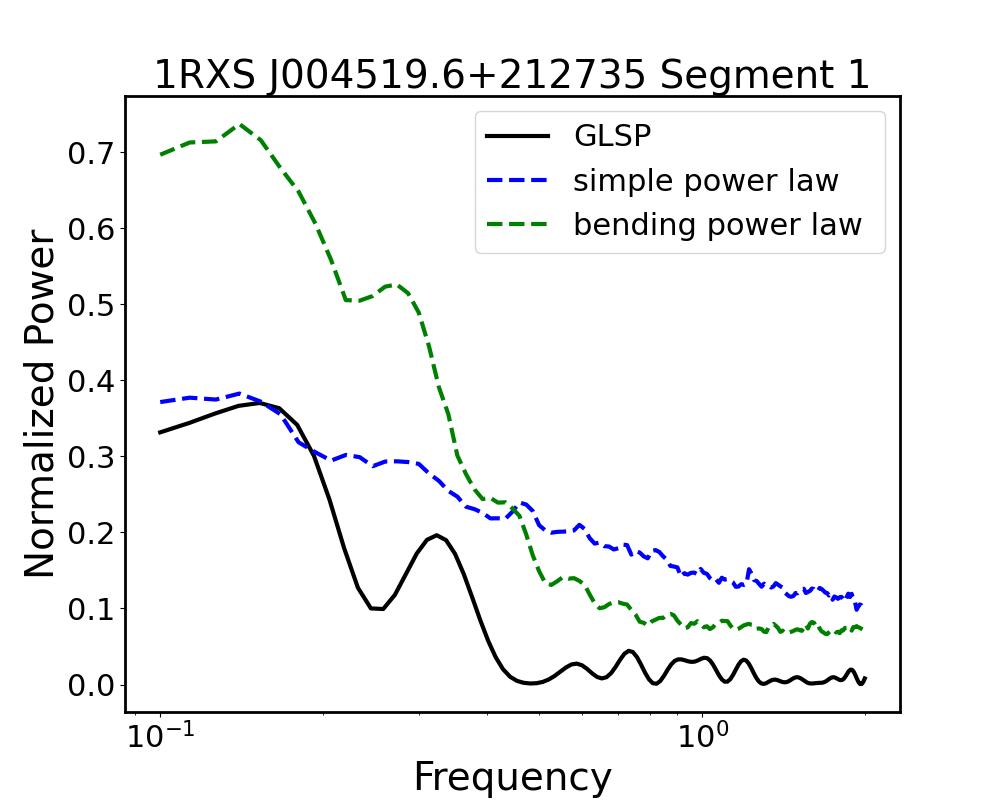}\includegraphics[scale = 0.25]{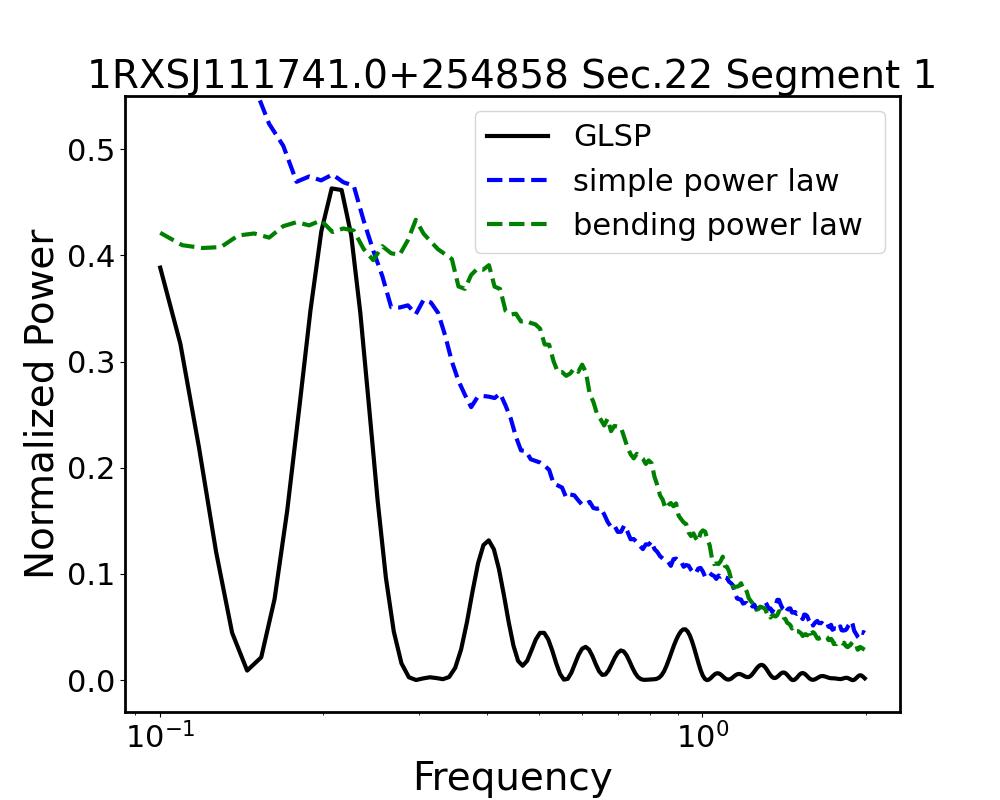}\includegraphics[scale = 0.25]{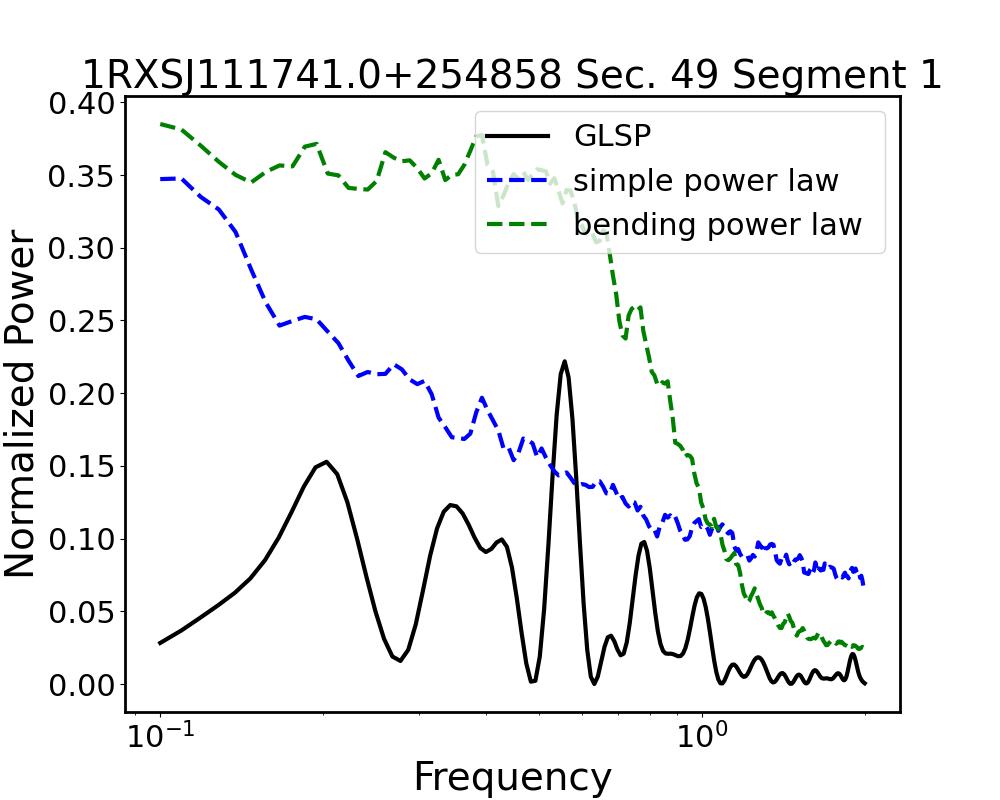}\\
\includegraphics[scale = 0.25]{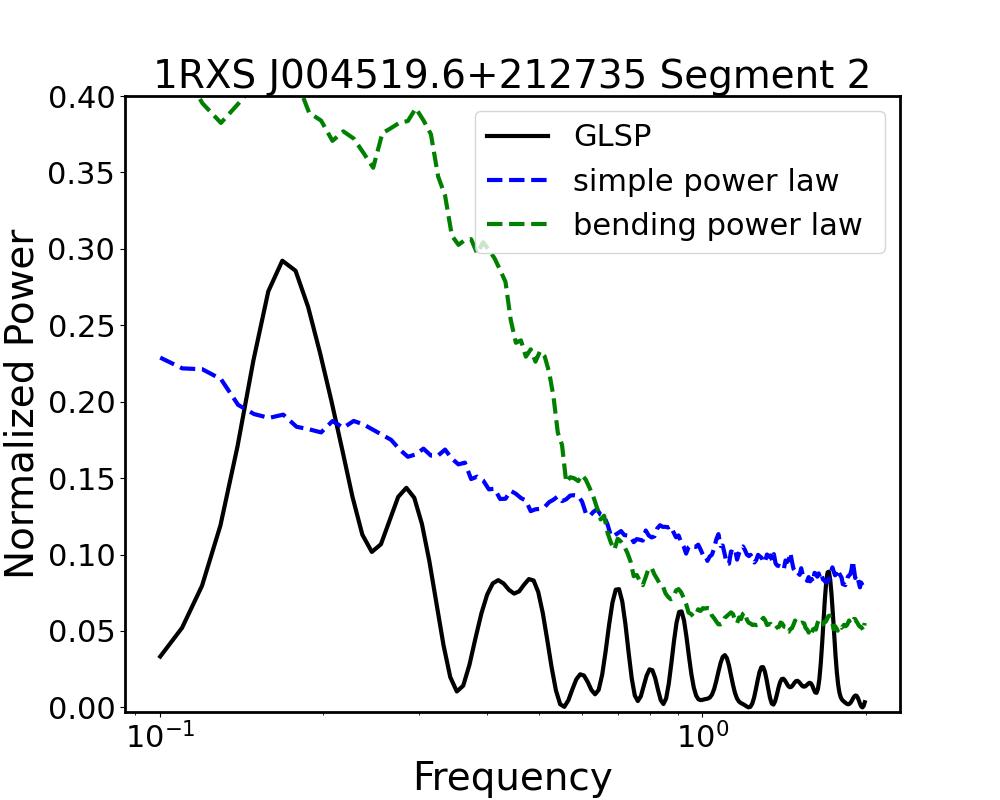}\includegraphics[scale = 0.25]{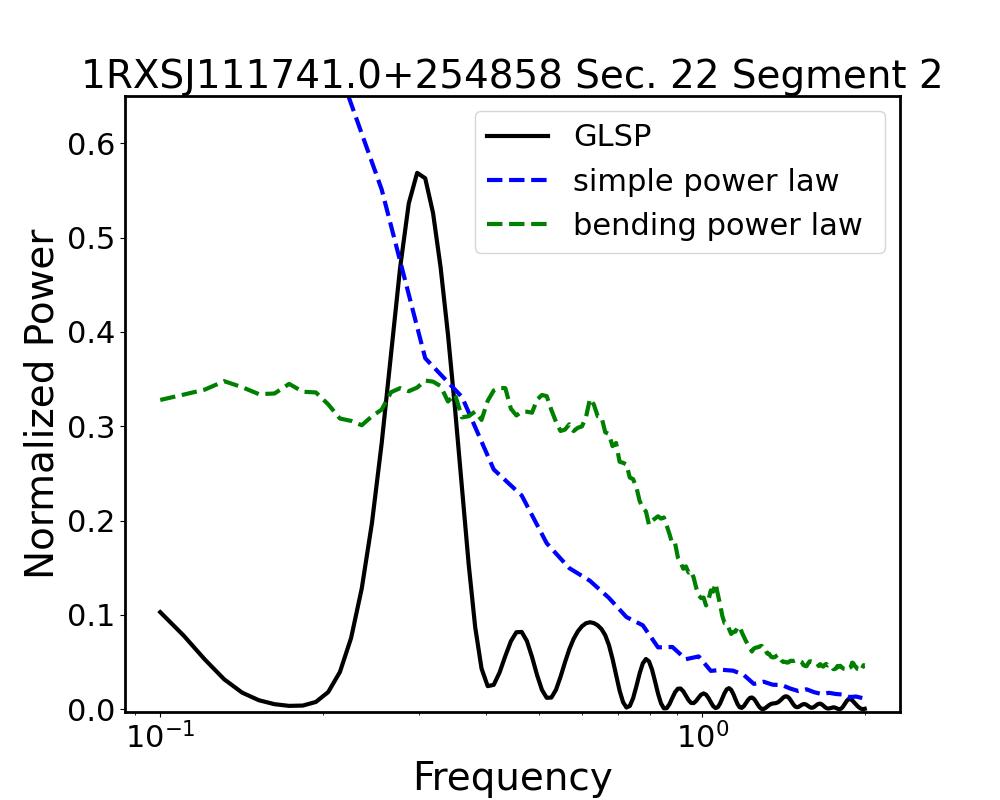}\includegraphics[scale = 0.25]{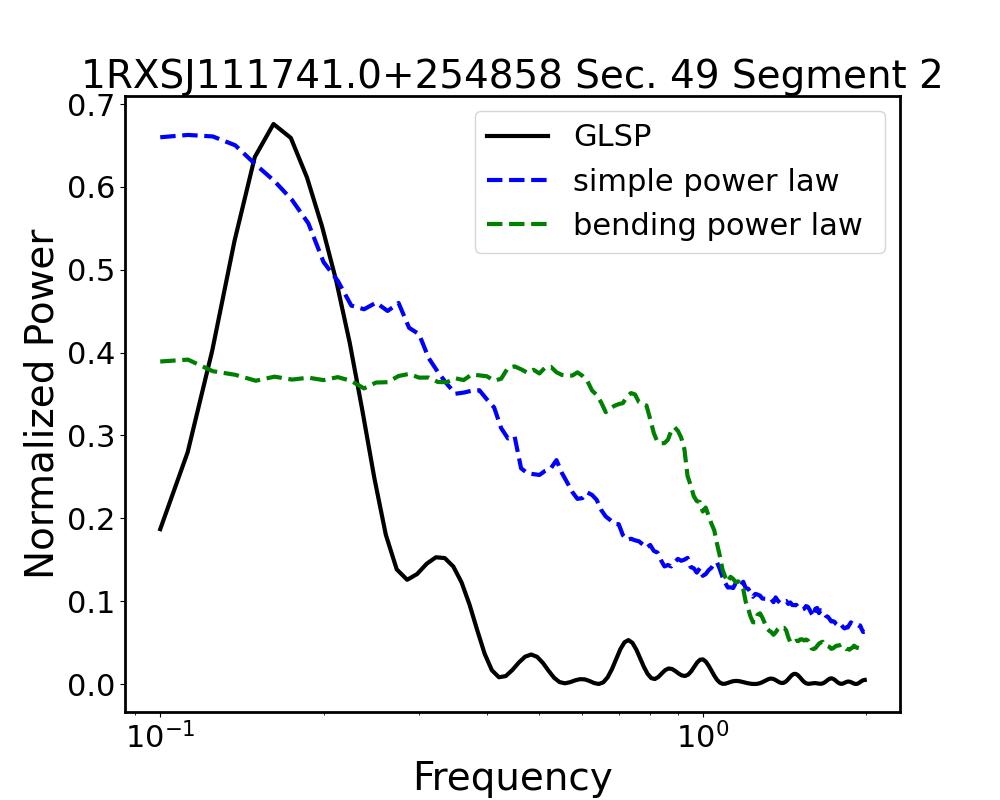}
\caption{Segment-wise GLSP analysis for the light curves of 1RXS J004519.6+212735 (left panel)  and 1RXS J111741.0+254858 (middle and right panels). In each plot, the black curve is the GLSP. The green and blue curves show the 3 $\sigma$ significance level for the simple power law and the bending power law, respectively.}
\end{figure*}\label{fig:seg}

The stochastic processes occurring in the jets and accretion disks of a blazar may be reflected in  observations taken throughout the whole electromagnetic spectrum. In general, they may be  non-stationary as their statistical properties can change with time. It is possible that the light curves of blazars may be non-stationary in nature even on the relatively short timescales investigated in this work. Indeed, evidence of non-stationarity has been observed in X-ray light curves of blazars and Seyferts with similar durations as the TESS light curves explored here \citep[e.g. ][]{2019MNRAS.485..260A,2020ApJ...897...25B}. However, the simulation procedure we have used to this point to generate significance curves assumes the simple or bending power-law description of the PSD is constant throughout the observation and therefore, should be considered cautiously. While the observations of BL Lac have only a short ($\sim 1$d) gap, the observations of 1RXS J004519.6+212735 and 1RXS J111741.0+254858 have substantial gaps. It is important to examine whether any QPO signal is present both before and after the gap or if it is only present in one of the segments. %of the light curves of these blazars.
To investigate  this issue, we have divided the light curve into two segments separated by a gap in the middle and performed the GLSP analysis assuming both simple and bending power laws as the underlying red noise model. The portion of each light curve before the 4-day gap is termed as Segment 1 and that after the gap is termed as Segment 2.

The left panels of Fig.\ 7 show the segment-wise analysis for the Sector 17 observation of 1RXS J004519.6+212735. In Segment 1, a peak at around 6.0 days is found with at least 3 $\sigma$ significance using the simple power-law but no peak is found significant employing a bending power-law.  For Segment 2, a similar signal  at around 5.9 days is significant at $>3 \sigma$  using the simple power law. While these peaks are not found to be significant using the bending power law model, their GLSP significance increases to $3 \sigma$ when  both segments are considered together (see Fig.\ 3) and is further supported by the wavelet analysis (Fig.\ 4).
%***THIS DOES NOT SEEM NECESSARY AND COULD BE CONFUSING***The reason could be the presence of high white noise level at higher frequencies for the individual segments which gets minimized when analyzing both segments together and increases the power of peak at lower frequencies. The normalized power of both individual and combined segment analysis has similar levels (normalized power $\approx$ 0.3). The difference is the significance level which is altered due to the increase in white noise level. In both segments, a period of around 6 days is found, which is also confirmed in the wavelet analysis (upper right panel of Fig.\ 4) of both segments combined.     ***BUT THIS LONG PERIOD HAS TOO FEW CYCLES TO BE CONSIDERED REAL, EVEN FOR THE COMBINED DATA***

We analyzed the Sec.\ 22 observation of 1RXS J111741.0+254858 by segments and the GLSP results are shown in the middle panels of Fig. \ 7. Signals of at least 3 $\sigma$ significance are found at  4.5 days for Segment 1 and 2.8 days for Segment 2, assuming either simple or bending power laws as the underlying model. In the combined analysis (Fig.\ 3), these two peaks blend into a double-horned peak.  This transition from one frequency to another is also evident in the color-color wavelet analysis of Sec.\ 22 (Fig.\ 4).

For the Sec.\ 49 observation of 1RXS J111741.0+254858, the segment-wise GLSP results are plotted in the right panels of Fig.\ 7. In Segment 1, a period of around 1.8 days is found to be significant at 3 $\sigma$  with the simple power law, though not with the bending one. This signal is also seen in the combined wavelet analysis of this observation (Fig.\ 4) but with less significance, as it is relatively strong only in the first half of the observation. In Segment 2, a very strong signal with at least 3 $\sigma$ with both simple and bending power law is found at a period of around 6 days. This evolution of the possible $\sim$6 day period is shown in the combined wavelet analysis (Fig.\ 4) where the power of the signal increases with time. %So, the periodicity claimed for this observation is 6.2 days which essentially comes from a very strong signal in the second segment of this observation.
%A signal of around 3 days with less significance is present in both segments and also confirmed in the combined WWZ analysis.  ***AGAIN THE 6 DAY PERIOD IS TOO LONG*** }

%This behavior is not necessarily unexpected in the context of jet instabilities.

\section{Discussion}\label{sec:con}

\begin{figure*}
\centering
\includegraphics[scale = 0.3]{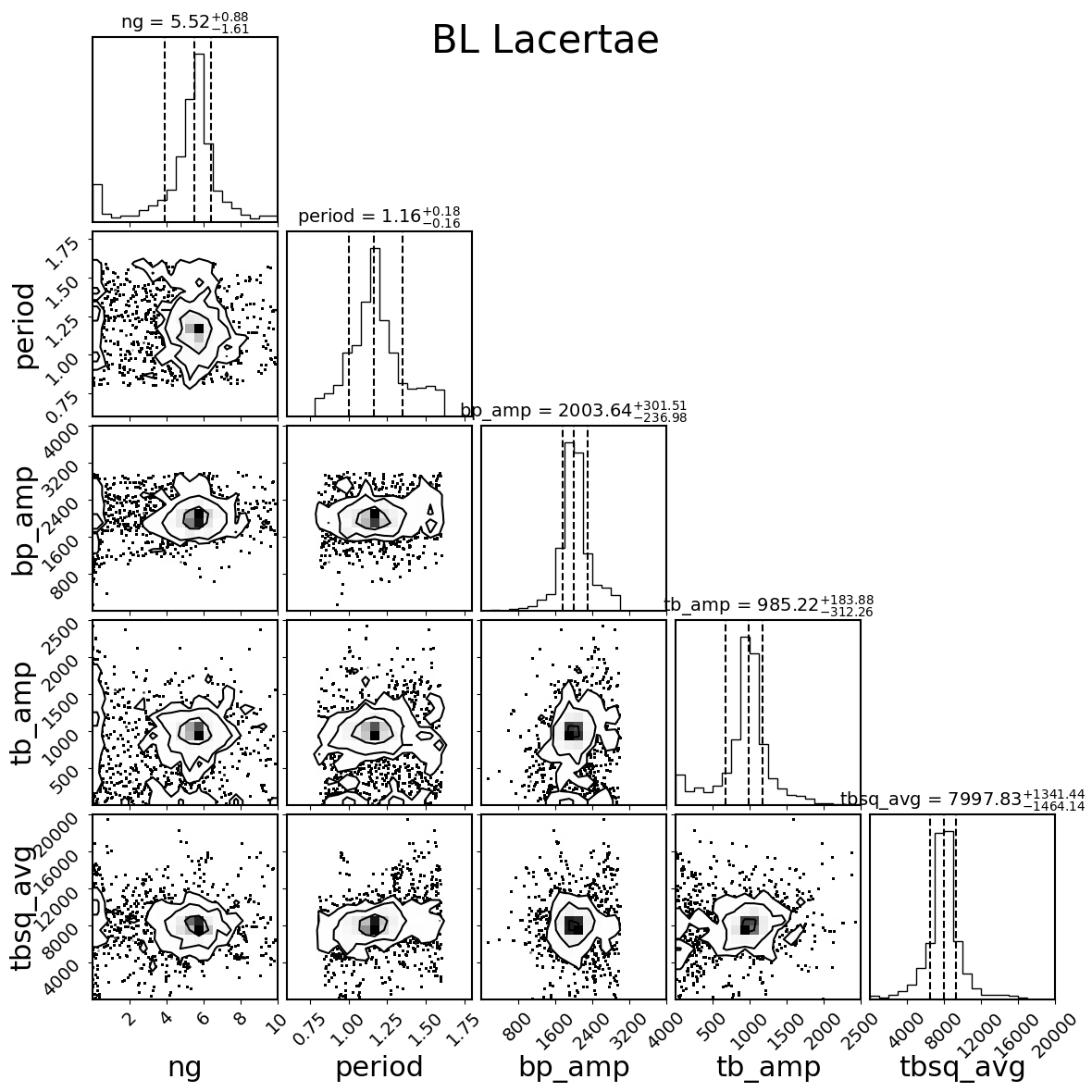}\includegraphics[scale = 0.3]{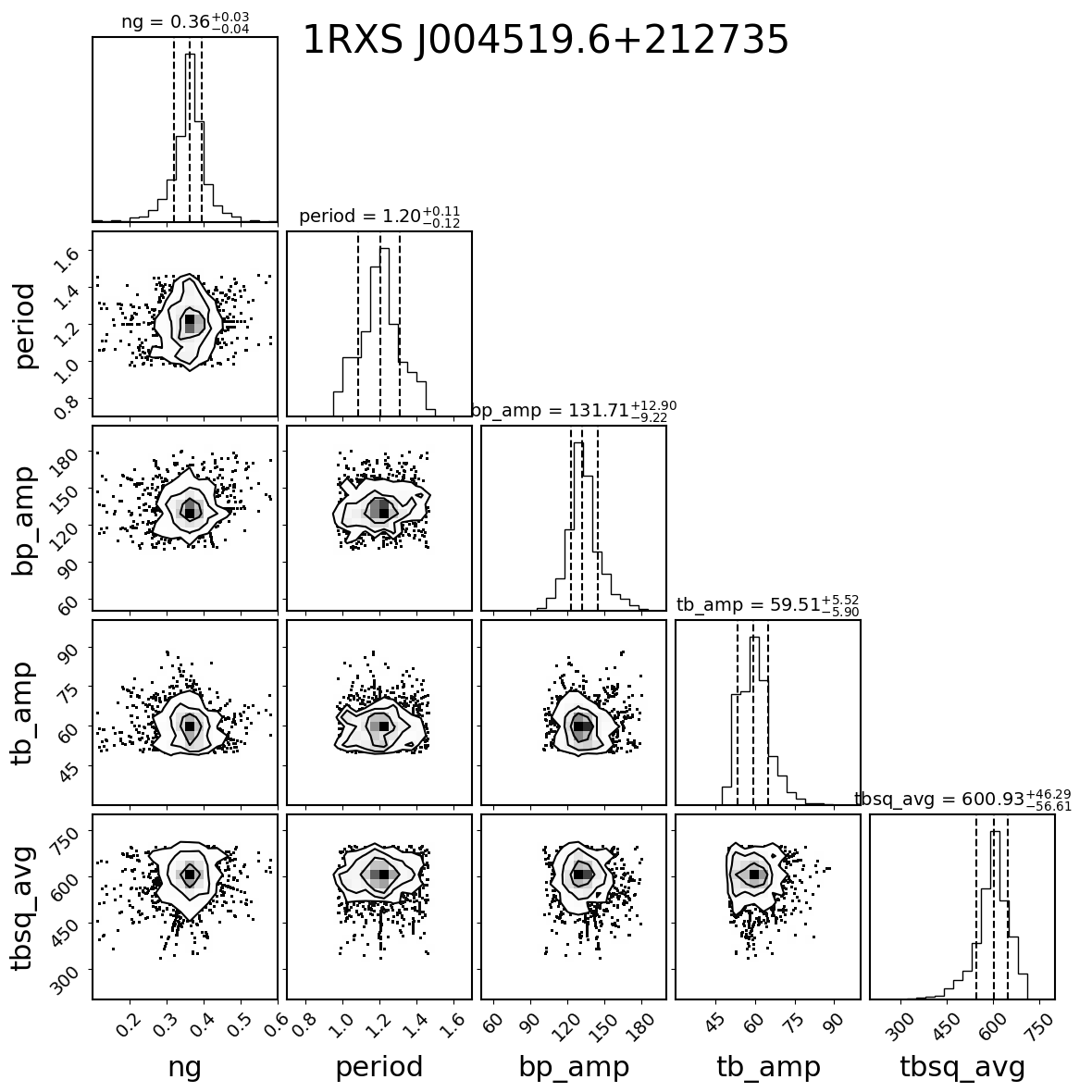}
\caption{Corner plot of posterior distributions sampled from 100 walkers and 120000 iterations through MCMC simulations assuming the kink instability model for BL Lacertae (left) and 1RXS J004519.6+212735 (right).}
\end{figure*}\label{fig:mcmc16}

\begin{figure*}
\centering
\includegraphics[scale = 0.3]{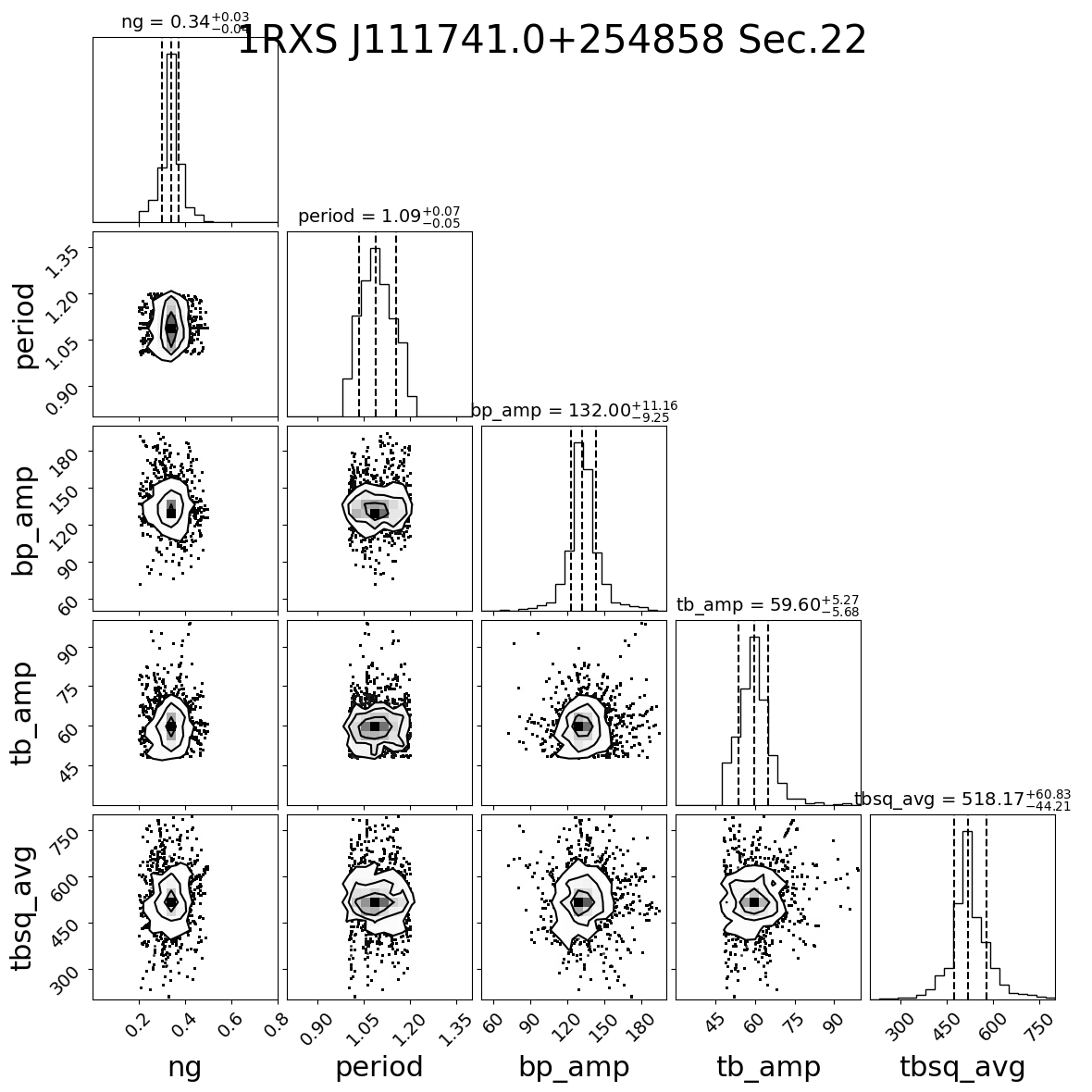}\includegraphics[scale = 0.3]{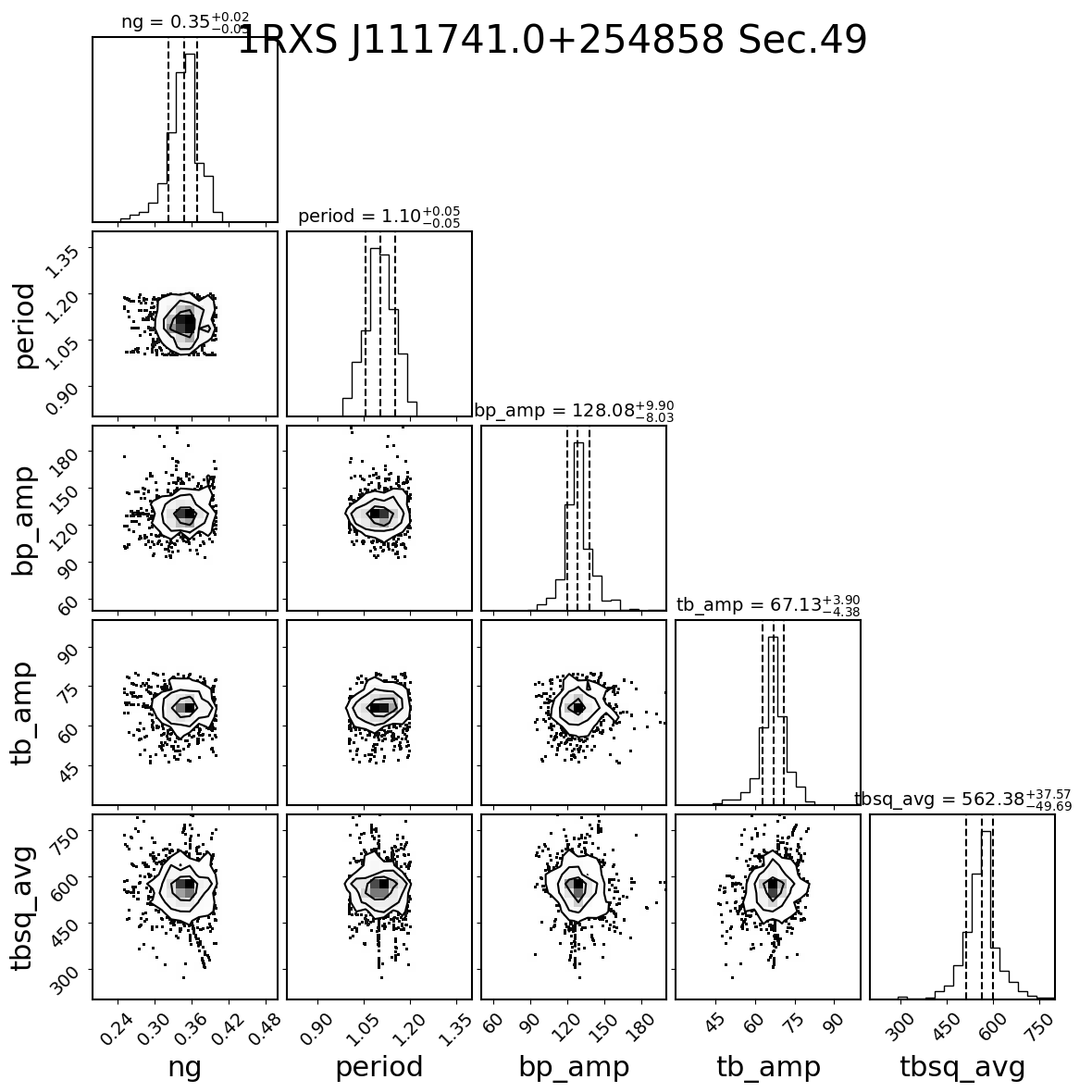}
\caption{Same as Fig.\ 8 but  %fig.~\ref{fig:mcmc16}
 for Cycle 2 (left) and Cycle 4 (right) observation of 1RXS J111741.0+254858.}
\end{figure*}\label{fig:mcmc22}

Emission from blazars is believed to be dominated by non-thermal processes in the relativistic jet. As the jet is viewed at a very small inclination angle ($<10^{\circ}$) \citep{1995PASP..107..803U}, the emission from the jet is magnified enormously by the relativistic effects and outweighs the cumulative emission from the host galaxy and accretion disk \citep[e.g.][]{2001ApJS..134..181J}. The emission, extending from radio to $\gamma$-rays, is observed to have variability on timescales of a few minutes \citep[e.g.][]{2017A&A...603A..25A, 2016ApJ...824L..20A, 2022Natur.609..265J} to years \citep[e.g.][]{2010MNRAS.402.2087V, 2014APh....54....1A}. Some variability timescales have also been claimed to be QPO signatures in blazars \citep{1998ApJ...504L..71H, 2013MNRAS.436L.114K,  2015Natur.518...74G, 2021MNRAS.501.5997T, 2022Natur.609..265J}. The short timescales of much of the variability indicate that the emitting region must be compact.

\subsection{Kink Instability}

Very recently, \cite{2022Natur.609..265J} reported a quasi-periodic feature with a 0.57 day period in BL Lacertae, one of the objects analyzed in this work, employing both optical and $\gamma$-ray observations that were taken between March and December 2020. They considered a kink instability as the reason for the origin of these QPOs based on the approach given by \cite{2020MNRAS.494.1817D}. In this model, periodicity is the result of a type of current-driven plasma instabilities known as kink instabilities. Relativistic magnetohydrodynamic simulations have shown that these kinds of instabilities can arise in the jet having a strong toroidal magnetic field and result in both twisting of field lines and an increase in particle acceleration \citep{2009ApJ...700..684M,2017MNRAS.469.4957B}. This distorted magnetic field grows to produce a quasi-periodic  kink in the jet that manifests into a region of compressed moving plasma with enhanced emission that could be observed to have a QPO signature. The timescales of such QPO signatures in blazars can  be of the order of days and are related to the growth of kink instabilities over timescales of weeks to months.

\cite{2022Natur.609..265J} performed semi-analytical  simulations to model the time-dependent emission from the plasma following \cite{2020MNRAS.494.1817D}. This model assumes the dominant source of fluctuations arises from a blob in the kink region and has three components: a constant toroidal magnetic component; a sinusoidal poloidal component having temporal period $T$ and amplitude $bp_{amp}$; and a  turbulence profile having amplitude $tb_{amp}$ which averages to $tbsq_{avg}$; $n_g$ corresponds to the average emission power during the observation. They constrained these parameters along with the normalization using MCMC simulations and found the overall magnetic (and hence polarization) period to be 1.14 days, which corresponds to the observed flux variability QPO of 0.57 d. They attributed the 4-day QPO period also detected using the wavelet method to the growth of this one-day period kink with time.

Quantifying the transverse motion of the kinked region can be used to calculate its growth rate \citep{2009ApJ...700..684M}. \cite{2020MNRAS.494.1817D} estimated the kink growth  time ($\tau_{KI}$) to be the ratio of the transverse displacement of the jet from its center ($R_{KI}$) to the average velocity of propagation ($\left< v_{tr} \right>$). They found the growth time of the kink instability to be consistent with the period of the QPOs estimated in their simulations. The period associated with the kink instability ($T_{obs}$) in the observer's frame is given by \citep{2020MNRAS.494.1817D}
\begin{equation}
T_{obs} = \frac{ R_{KI} }{ \left< v_{tr} \right>\delta }~,
\end{equation}
where $\delta$ is the Doppler factor and $R_{KI}$ is the size of the emission region in the co-moving frame and its value can be $10^{16}$--$10^{17}$ cm for a typical BL Lac object. In \citet{2020MNRAS.494.1817D}, $\left< v_{tr} \right>$ is estimated to be $\approx 0.16c$. So, the kink instability is related to the properties of the jet which include its inclination, velocity, hence the Doppler factor, and size of the emission region. Calculating $T_{obs}$ for BL Lacertae using $\delta=15$ yields a time of 1--10 days which is consistent with the QPOs timescales found in this work. So it is reasonable to suggest that the apparent quasi-periods of 3--5 days we obtained here for all three sources using different timing analyses can be attributed to the growth of the quasi-periodic kinks originating in the inner portions of jets.

The longer QPO periods of around 3--5 days, indicated by both GLSP and wavelet approaches could be explained by the kink evolving with time. Any quasi-periods of more than $\sim$ 5 days cannot be firmly detected with the \textsl{TESS} light curves having a duration of 25--27 days as a period of more than 6 days corresponds to fewer than 4 cycles.
 It should also be noted that any QPOs of periods shorter than a few hours could not be reliably detected in these data sets as the sampling frequency of these \textsl{TESS} data is 30 minutes so they would be very close to the Nyquist frequency and lie in the white-noise portion of the power spectrum.

Still, there are  other processes that might be able to explain any QPOs on the scale of a few days. Possible explanations include the normal modes of oscillations trapped in the inner part of the accretion geometry due to the strong gravity of the central compact object \citep[e.g.,][]{1997ApJ...476..589P,2008ApJ...679..182E}. These modes correspond to the internal oscillations of the accretion disk and could produce the observed QPOs. Magnetorotational instability (MRI) turbulence in the accretion flow \citep[e.g.,][]{2004ApJ...609L..63A} might also result in such periodicity. The frame-dragging around a rotating SMBH leads to the Lens-Thirring precession near the inner portion of the accretion disk which could naturally produce a QPO with a period of the order of a few weeks \citep{1998ApJ...492L..59S}. However, these processes are more likely to be more important in Seyfert galaxies and quasars where the emission from the disk dominates over that of the jets, whereas the jet emission is the most important in these blazars.

\subsection{Individual Sources}
\subsubsection{BL Lacertae}

BL Lacertae is the prototype of the BL Lac objects characterized by very weak or absent broad emission lines and high variability observed throughout the electromagnetic spectrum. This object has been observed extensively in all wavebands, from radio to $\gamma$-rays \citep[e.g.][and references therein]{2018A&A...615A.118S,2003A&A...402..151R,2023A&A...672A..86R}.
It also has been claimed to show quasi-periodic behavior in various wavebands with periods ranging from a few minutes to hours, weeks, and even years. \citet{2004A&A...424..497V} claimed a periodicity of 8 years for radio outbursts for this source which was later supported by \citet{2009A&A...501..455V}. \citet{2017A&A...600A.132S} reported a periodicity of around 700 days in both optical and $\gamma$-ray observations. As mentioned above, \cite{2022Natur.609..265J} analyzed the optical and $\gamma$-ray observations and claimed a periodicity of around 13 hours in both. They found a very strong correlation between the optical and $\gamma$-ray light curves (correlation coefficient: 0.62$\pm$0.04) with a temporal lag between them  consistent with 0. % not significantly significant ($0.02^{+0.05}_{-0.44}$). Besides, both the light curves show a periodicity of 0.55 days.
The detection of essentially identical periodicites implies that both optical and $\gamma$-rays are produced from the same mechanism which is claimed to be the kink instability in the jet. They also found a periodicity of the order of a few days which is explained by the temporal growth of the kink. %As the non-thermal emission from the blazars is dominated by the jet emission and can be observed throughout the electromagnetic spectrum, these optical and $\gamma$-ray emissions originate in the jet with the same emission mechanism.

In our work, the periodicity of $\sim$5 days is found in the \textsl{TESS} light curves of BL Lacertae, which we suggest is the optical counterpart of the kink developed in the jet.
BL Lacertae was observed by \textsl{TESS} for 27~days beginning on 12 September 2019,   approximately 5 months prior to the onset of the observations analyzed by \citet{2022Natur.609..265J}. \citet{2023A&A...672A..86R} analyzed the weekly and monthly binned Fermi-LAT observations of BL Lacertae taken between August 2008 and April 2021 and found no significant long-lived periodicity.  Although the \textsl{TESS} observation analyzed in this work is subsumed within the span of these Fermi observations, the cadences are  significantly different. The minimum cadence normally available with Fermi data is 1 day and they usually need to be binned over longer periods to get adequate statistics.  Therefore, the Fermi observations with a 1 day cadence could not resolve periodic variability with timescales of just a few days, while the 30 minute cadence of the Cycle 2 TESS data can.

We have applied the kink instability model to the \textsl{TESS} observation of BL Lacerate and followed the approach of \citet{{2022Natur.609..265J}} to fit our 26-day observation using their publicly available
code\footnote{\url{https://zenodo.org/record/6562290}}. Fig. % ~\ref{fig:mcmc16}
8 displays the corner plots showing the posterior distributions of various parameters using the algorithm applied in the code. We used the Doppler factor $\delta = 15$ which is derived in \cite{2001ApJS..134..181J} for  BL Lacertae using Very Large Baseline Array (VLBA) observations at 43 GHz. We used 100 walkers with 120000 iterations and fit 5 parameters used in the code. We obtained the period of 1.16 days which agrees nicely with the results of \cite{2022Natur.609..265J}.

 \subsubsection{1RXS J111741.0+254858 and 1RXS J004519.6+212735}

1RXS J111741.0+254858 \citep{2010A&A...518A..10V,2020yCat.1350....0G} and 1RXS J004519.6+212735 \citep{2007ApJS..173..471M,2020yCat.1350....0G} were found serendipitously in the  SDSS-IV SPIDERS catalog, which is essentially a catalog of X-ray-selected Seyferts. For 1RXS J111741.0+254858, we have analyzed the observations from Sec.\ 22 (Cycle 2) and Sec.\ 49 (Cycle 4) and found probable QPOs of 3.7 and 4.6 days respectively. A QPO signal at 5.8 days is found in the analysis of the Sector 17 observation of 1RXS J004519.6+212735.

Motivated by the successful application of the kink instability model to BL Lac's \textsl{TESS} data, we applied this formalism to these other two sources.  For 1RXS J004519.6+212735, the period is found to be 1.20 days. For the Cycle 2 and Cycle 4 observations  of 1RXS J111741.0+254858, periods of 1.09 and 1.1 days are found, respectively.  The corresponding MCMC corner plots are given in Figs.\ 8--9. For this analysis, we used the Doppler factor appropriate for BL Lac itself, $\delta = 15$ for both sources. We also examined whether using different values of $\delta$ would yield different temporal periods. Hence we also used the average Doppler factor value of 9.2 derived in \cite{2017MNRAS.466.4625L} for BL Lac objects. For 1RXS J004519.6+212735, the new period is found to be 1.26 days, while periods of 1.19 and 1.21 days are found for the Cycle 2 and Cycle 4 observations of 1RXS J111741.0+254858, respectively. So, plausible variations in $\delta$ have only modest effects on the observed period of the kink developed in the jet.

\section{Conclusion}
In this work, we have examined densely sampled optical light curves  of three blazars, BL Lacertae, 1RXS J111741.0+254858, and 1RXS J111741.0+254858.  These were extracted from \textsl{TESS} using a customized reduction approach designed to preserve actual AGN variability. We have discovered tentative evidence for QPOs in these light curves on the timescale of a few days. For each light curve, these possible QPOs mainly lie between $\sim 2$ and $\sim 6$ days. We also ran MCMC simulations to examine the claimed QPOs present in the observations, assuming they originate with a kink instability in the jet.  We found the periods of the kinks to be consistent with the period reported for BL Lacertae in \citet{2022Natur.609..265J}. %The periodicity manifested in the GLSP and WWZ analysis in this work is explained by the growth of kink with time.
Our analysis thus supports the kink instability hypothesis where the apparently observed QPOs can be produced through the temporal evolution of a kink developed in the jet.

%\textcolor{blue}{Magnetic acoustic waves propagates within the corona and induces oscillation within corona\citep{2010MNRAS.404..738C, 2016AN....337..398M}. This attenuates the efficiency of Comptonization process which produces periodicity.}
%{\color{red} \\

\section*{Acknowledgements}
 This work is supported by NASA grant number 80NSSC22K0741. KLS gratefully acknowledges the staff of the K2 Guest Observer office at NASA Ames, especially Christina Hedges, for their assistance and advice in adapting the matrix regression methods.

%{\it Facilities}: Transiting Exoplanet Survey Satellite (TESS)

%{\it Software}: {\tt TESSCut} \citep{2019ASPC..523..397B}, {\tt Lightkurve} \citep{lightkurve}, {\tt Astropy}\citep{2022ApJ...935..167A}, and {\tt stingray}\citep{2019ApJ...881...39H}.
\section*{Data Availability}
The TESS data presented in this paper were obtained from the Mikulski Archive for Space Telescopes (MAST) at the Space Telescope Science Institute.

\appendix

{}

\end{document}